\documentclass[12pt, preprint]{aastex}

\usepackage{amsmath}
\usepackage[dvips]{color}

\newcommand{\sax}{SAX J1712.6-3739 }

\shorttitle{Microquasar bow shock nebulae}

\shortauthors{Yoon et al.}

\begin{document}

\title{Jet trails and Mach cones: The interaction of microquasars with
  the ISM} 

\author{
Yoon, D.\altaffilmark{1}, 
Morsony, B.\altaffilmark{1}, 
Heinz, S.\altaffilmark{1}, 
Wiersema, K.\altaffilmark{2}, 
Fender, R.P.\altaffilmark{3},
Russell, D.M.\altaffilmark{4}, \&
Sunyaev, R.\altaffilmark{5,6}  
}

\altaffiltext{1}{Department of Astronomy, University of Wisconsin-Madison, Madison, WI, USA}
\altaffiltext{2}{University of Leicester, University Road, Leicester LE1 7RH, UK}
\altaffiltext{3}{School of Physics \& Astronomy, University of Southampton, Highfield, Southampton, S017 1BJ, UK}
\altaffiltext{4}{Astronomical Institute ‘Anton Pannekoek’, University of Amsterdam, P.O. Box 94249, 1090 GE Amsterdam, the Netherlands}
\altaffiltext{5}{Space Research Institute (IKI), Moscow, Russia}
\altaffiltext{6}{Max-Plank-Institut f\"ur Astrophysik, Garching, Germany}

\begin{abstract}
  
  A sub-set of microquasars exhibit high peculiar velocity with
  respect to the local standard of rest due to the kicks they receive
  when being born in supernovae.  The interaction between the radio
  plasma released by microquasar jets from such high-velocity binaries
  with the ISM must lead to the production of trails and bow shocks
  similar to what is observed in narrow-angle tailed radio galaxies
  and pulsar wind nebulae.  We present a set of numerical simulations
  of this interaction that illuminate the long term dynamical
  evolution and the observational properties of these microquasar bow
  shock nebulae and trails.  We find that this interaction always
  produces a structure that consists of a bow shock, a trailing
  neck, and an expanding bubble. 
  Using our simulations to model emission, we predict that the shock 
  surrounding the bubble and the neck should be visible in H$_\alpha$ 
  emission, the interior of the bubble should be visible in synchrotron 
  radio emission, and only the bow shock is likely to be detectable in 
  X-ray emission.
  We construct an analytic model for the evolution of the neck and 
  bubble shape and compare this model with observations of X-ray binary
  SAX J1712.6-3739.

\end{abstract}
\keywords{x-ray binaries, microquasars, jets}

\section{Introduction}
It is now well established that accretion onto black holes can lead to
the production of powerful jets, both in the case of AGN, in which
case we call the object a radio galaxy, and in the case of X-ray
binaries (XRBs), in which case we call the object a microquasar.  We
have since learned that even neutron star X-ray binaries 
as well as some accreting white dwarfs behave in the
same way \citep{fen04,mig06,tud09,kording08}.

When powerful jets run into their environment, they produce bright
external shocks (called hot spots or working surfaces) and generally
inflate pockets of relativistic plasma that emit diffuse synchrotron
emission.  If the black hole is stationary, these pockets take on
roughly ellipsoidal shapes and are called radio lobes or cocoons. They
are surrounded by the interstellar or intergalactic gas that was
occupying the volume prior to inflation, compressed into a narrow
shell. This picture is most easily observed in the case of X-ray
cavities in galaxy clusters inflated by AGN jets \citep[][and
references therein]{rafferty:06}, but has also been discovered around
a few microquasars \citep{gallo05,hao09}.

However, when the black hole is moving at appreciable speed with
respect to its environment, the ram pressure of the headwind can
dramatically alter the dynamics of the outflowing radio plasma,
sweeping it back and generating a bow shock ahead of the moving black
hole.  This has long been known in the case of radio galaxies
\citep[e.g.][]{miley72}, which, depending on the angle at which the
plasma is bent back, are called narrow or wide angle tail sources
(generically, these sources are also called ``bent doubles'').

Given that the black holes at the core of microquasars are born in
supernova explosions, and given that supernovae can impose significant
kick velocities on the compact objects they produce, \cite{hei08}
suggested that a similar phenomenon to ``bent doubles'' should occur around a
sub-population of microquasars: the high-velocity tail of the
population of low-mass X-ray binaries (LMXBs).  It is now known that
several microquasars are moving through the ISM at relative velocities
in excess of $v_{\rm ext}>100$ ~km s$^{-1}$ \citep{mir01}.  

The phenomenology of these trailed microquasars should be broadly
similar to their supermassive AGN equivalent, leading to the
production of a low-surface brightness trail of synchrotron emitting
relativistic plasma, and a brighter bow shock nebula. While the mode
of inflation is fundamentally different, the end product should also
be very similar to pulsar wind bow-shock nebulae
that are formed by interaction between strong wind from pulsars with 
significant kick velocities in the interstellar medium. The general 
bow shock features have been well observed in H$_\alpha$ emission for 
PSR B1957+20 \citep{stappers:03}, PSR B0740-28 \citep{stappers:02}, 
PSR J0437-4715 \citep{bell:95}, and PSR B2224+65 (the ``guitar'' nebula) \citep{cha02}. 
Also, G359.23-0.82 which is powered by PSR J1747-2958 shows bright head 
and X-ray \& radio ``trails'' \citep{gaensler:04}.

The first candidate of a bow shock around such a trailed source, the
LMXB \sax, was discovered in an H$_\alpha$ image by \citet[][originally
found by \citealt{int99}]{wie09}.  It is broadly consistent with the
predictions of \cite{hei08}.  Future searches for large scale nebula
and diffuse synchrotron emission around known LMXBs are needed to
further test the predicted existence of this population of sources.

The model developed in \cite{hei08} was purely analytic and a number
of important aspects of the evolution of such trailed microquasars was
left open.  Numerical simulations are needed to investigate the
production of these sources in more detail and to confirm the
qualitative predictions of \cite{hei08}.

In this paper, we model the evolution of the XRBs moving through the
ISM by using 3-dimensional hydrodynamic simulations and to derive a
more detailed dynamical model for the large scale evolution of these
sources.

A key aim of this study is to develop quantitative diagnostics that
can be used to derive important constraints on the core parameters of
microquasars from observations of trailed microquasars, such as their
age, their relative velocity with respect to the local standard of
rest ($v_{\rm XRB}$), and the jet power from observational parameters like
the opening angle of the bow shocks, the size of the hot bubble at the
terminus of the trail, and the brightness of the shock, trail, and
bubble.  In addition, our simulation results can be used to predict
the brightness of H$_\alpha$, bremsstrahlung, X-ray, and radio
synchrotron emission and to design targeted observational searches for
these sources.

This paper is organized as follows: In \S 2, we present the numerical
method and the initial conditions for our numerical study.  In \S 3, we
discuss the evolution of XRBs, scaling relations, and observational
expectations. In \S 4, we compare our results with new observations of
the LMXB SAX J1712.6-3739. Finally, in \S 5 we summarize
our results.

\section{Technical Description}
\subsection{The code}
Simulations are carried out in 3 dimensions with the FLASH 2.4
hydrodynamic code \citep{fry00}, which is a modular, adaptive mesh
refinement code.  It solves the Riemann problem using the
piecewise-parabolic method.  The code is formally accurate to second
order.  The gas is modeled with either an adiabatic equation of state
with index $\gamma$=5/3 or with radiative line cooling.  Radiative
cooling from an optically thin plasma is implemented in FLASH adopting
a piecewise-power law approximation that evaluates a plausible fit to
$\Lambda(T)$ within the temperature range of $4.4 \times 10^{3} < T <
10^{8} \textrm{K}$

\subsection{The jet nozzle}
In order to simulate the injection of collimated, supersonic jets into
the grid, we employ a numerical ``nozzle'', as first developed and
described in \citep{heinz06}: An internal inflow boundary of
cylindrical shape placed at the location of the XRB, injecting fluid
with prescribed energy, mass, and momentum flux to match the
parameters we chose for the jet.

For reasons of numerical stability, we impose a slow lateral outflow
with low mass flux in order to avoid complete evacuation of zones
immediately adjacent to the nozzle due to the large velocity
divergence at the nozzle.  The injection of energy and mass due to
this correction is negligible.

We generally follow the prescription for jet injection used in
previous simulations of AGN jets described in \citep{heinz06}, but
keep the location of the XRB fixed in space, instead letting the
external medium stream by at velocity $-v_{\rm ext}$.  Unlike our AGN
simulations, we do not impose a random jitter on the jet axis in this
case.

We chose to inject the jet at an internal Mach number of 10.  While
our adiabatic simulations are strictly scale free, we picked a set of fiducial
dimensions for our box that lead to the following natural scaling: For
computational feasibility, we chose a jet velocity of $v_{\rm
  jet}=3\times 10^{9}\,{\rm cm\,s^{-1}}$.  The jet is turned on
initially and continues to inject material for the entire length of
the simulation.

The simulations were carried out with Adaptive Mesh Refinement in
order to capture the large dynamic range required, ensuring that the
nozzle is resolved with at least 8 cells across.  For our fiducial
scaling, the maximum resolution for the standard model is about 0.156
pc near the jet nozzle. 

Based on the estimated power of the jet in Cyg X-1 from
\citet{gallo05}, the jet power in our simulation is set to a constant
value of $W_{\rm jet} = 10^{37} \textrm{ ergs s}^{-1}$ in our fiducial
scaling.  Note that since the Cyg X-1 is known to be one of the most
powerful XRB sources, sustaining the hard X-ray state for about 90\%
of the time.  The jets in our model might be more powerful than those
of typical LMXBs.  However, with the exception of our simulations with
cooling, our simulations are scale free and can thus be adjusted
easily to other parameter combinations.

\subsection{Initial conditions}

We placed the XRB in a moving medium inside a box large enough 
that boundary conditions never influence the dynamics.  We varied the 
velocity of the gas relative to the XRB to be $v_{\rm XRB} = 30$
~km s$^{-1}$, $100$~km s$^{-1}$ and $300$~km s$^{-1}$ (see Table \ref{tab:param}).  The
lowest velocity case represents a typical LMXB, given that the LMXB
velocity dispersion is $37\,{\rm km\,s^{-1}}$, while the largest
represents the most extreme case plausible (a marginally unbound
source).  The simulations we ran for this study are listed in
Tab.~\ref{tab:param}.

The fiducial ISM pressure we use is $P_0 = 3 \times 10^{-12} \textrm{ ergs
 cm}^{-3}$ following \citet{cox05}, and we use an ISM number density
of $n_{\rm ISM}=1 \textrm{ cm}^{-3}$, giving a sound speed in the ISM
of $c_s = 17.3$~km s$^{-1}$.

\begin{deluxetable}{ccccc}
  \tablecolumns{5} \tabletypesize{\scriptsize} \tablewidth{0pt}
  \tablecaption{Parameter of the simulations \label{tab:param}}
  \tablehead{ \colhead{Name} & \colhead{$v_{\rm XRB}$} & \colhead{Maximum
      Resolution} & \colhead{Luminosity} & \colhead{Radiative Cooling} } \startdata
  lv  & 30 ~km s$^{-1}$  & 0.312 pc & $10^{37}$~erg/s & no \\
  mv  & 100 ~km s$^{-1}$ & 0.156 pc & $10^{37}$~erg/s &  no \\
  hv  & 300 ~km s$^{-1}$ & 0.156 pc & $10^{37}$~erg/s &  no \\
  hvc  & 300 ~km s$^{-1}$ & 0.156 pc & $10^{37}$~erg/s &  yes \\ 
  fit1  & 100 ~km s$^{-1}$ & 0.039 pc & $8.47\times10^{33}$~erg/s &  yes \\ 
  fit2  & 200 ~km s$^{-1}$ & 0.039 pc & $8.47\times10^{33}$~erg/s &  yes \\ 
  fit3  & 300 ~km s$^{-1}$ & 0.039 pc & $2.29\times10^{35}$~erg/s &  yes 
\enddata
\end{deluxetable}

Typically, simulations were carried out to 1 Myr in our fiducial
scaling (much longer than the dynamical evolution of the jet, and long
enough for the quasi-steady state of the bow shock and the
self-similar solution we will discuss below to be established).  

Table \ref{tab:param} also includes one case with our standard
parameters and radiative cooling (with $v_{\rm XRB}$ = 300 ~km s$^{-1}$), the hvc
case.
We used this run to verify that the morphology of our simulations are
not strongly affected by cooling and to properly model emission (see
\S 3.3).  A further 3 simulations with cooling, fit1, fit2 and fit3,
were carried out to attempt to match H$_\alpha$ observations of SAX
J1712.6-3739 (see \S 4.1)

Because of the self-similar nature of the flow, the simulations lend
themselves to a staggered refinement scheme.  As we will describe
below, the radio plasma forms of a large scale bubble and a neck
connecting the XRB to this bubble. We can therefore de-refine the
simulation on cylinders on increasing radius around the axis of
propagation of the XRB and again de-refine with increasing distance
from the XRB along the axis of propagation.  We decrease the
refinement by a factor of two for every power of two increase in
radius and distance.

\section{Results}

\subsection{Morphological Evolution}

\cite{hei08} presented a simple, analytic model for the long term
dynamical evolution of trailed microquasars.  Our simulations allow us
to move beyond the initial heuristic model and describe the dynamics
of microquasar driven bow shock nebulae in detail.

The initial evolution of the system follows the prediction of
\cite{hei08} almost exactly: The jets inflate a single large,
roughly spherical bubble that expands following the self-similar
solution for a continuously driven bubble by \cite{castor:75},
surrounded by a thin shell of swept-up ISM.  

Initially, the expansion velocity of the bubble is much faster than
the space velocity of the XRB, $v_{\rm XRB}$, implying that the XRB
remains roughly at the bubble's center. As the expansion slows down,
the XRB begins moving towards the shell and eventually breaks out of
the expanding bubble.  

At this point, the dynamic pressure of the ISM due to the XRB's
velocity causes the jets to bend backwards, and a trail of radio plasma
is created behind the XRB, connecting it to the bubble.  The radio
plasma released by the XRB continues to inflate the bubble, akin to a
balloon inflated by a straw.  The bubble, therefore, continues to
expand spherically even though the binary can be far outside the
bubble.

As it propagates, the XRB is driving a bow shock into the ISM.  As
predicted in \cite{hei08}, dynamical instabilities develop at the
contact discontinuity between the bow shocked ISM and the backflow
along the radio trail, dissipating some of the backflow energy along
the channel.

\begin{figure}[!htbp]
   \centering
   \includegraphics[width=0.6\textwidth]{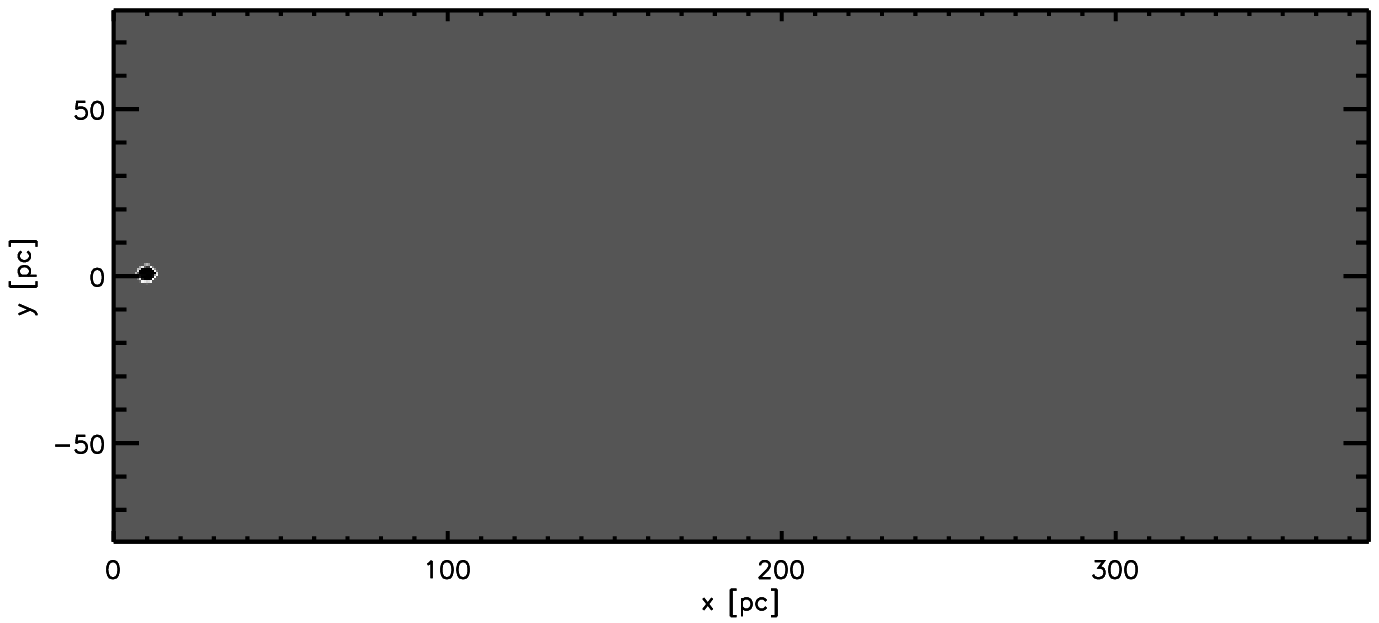}
   \includegraphics[width=0.6\textwidth]{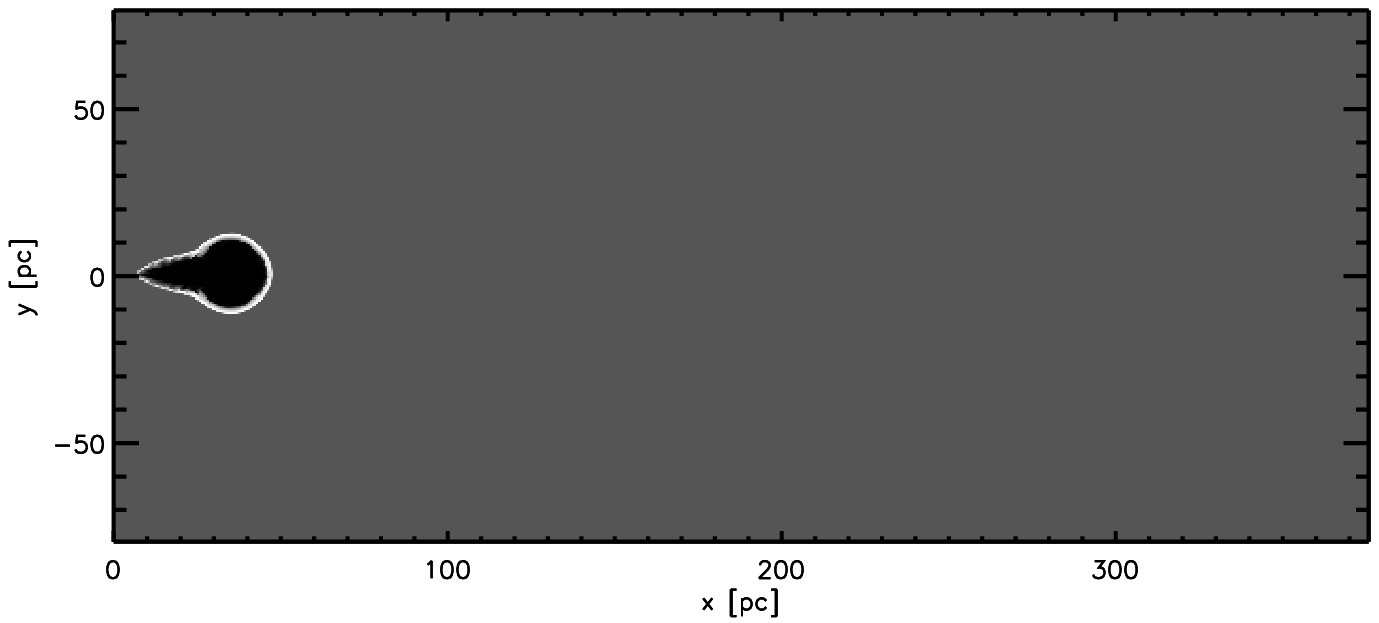}
   \includegraphics[width=0.6\textwidth]{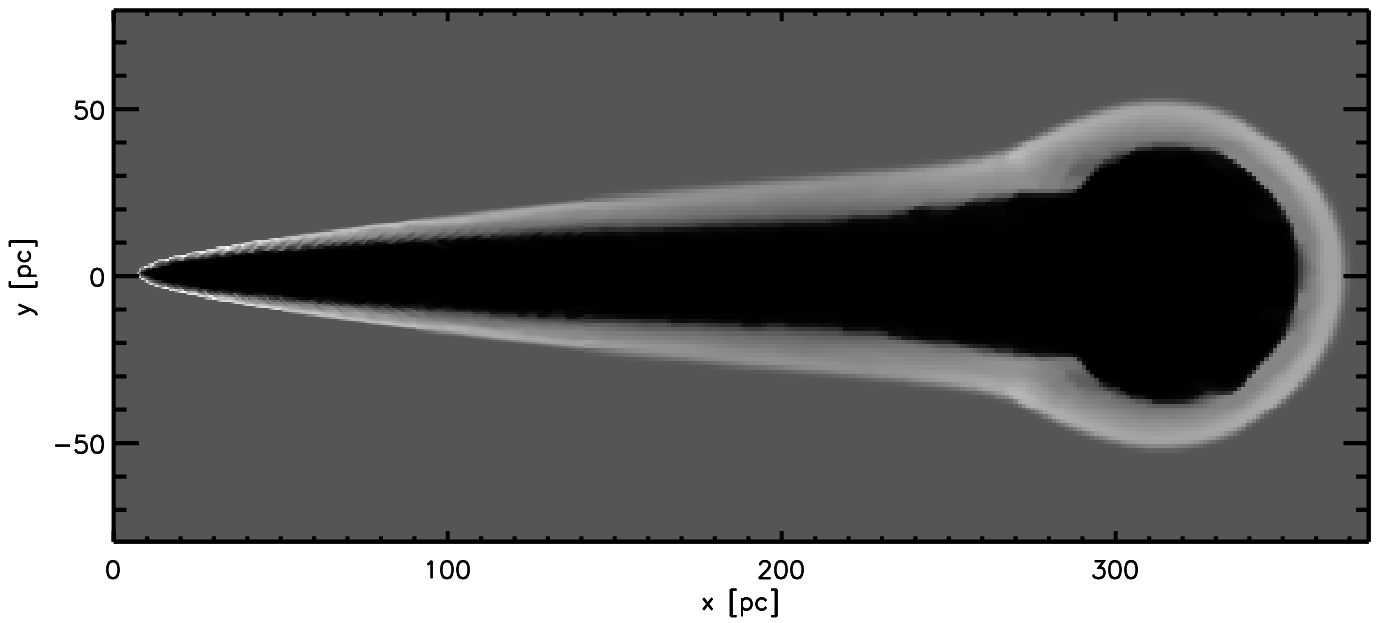}
   \caption{Density slice through our 300 ~km s$^{-1}$ simulation at 10,000
     years (top), 100,000 years (middle) and 1 Myr (bottom).  The XRB
     is located toward the left of the images.  At 10,000 years (top)
     the XRB is surrounded by a spherical bubble, although the XRB is
     nearing the left edge.  At 100,000 years (middle), the XRB has
     broken out of the bubble and created a distinctive shape.  The
     jets, directed up and down, are curved back at the bow shock
     (curved shock at left) and directed back to the large bubble to
     the right, where they terminate and power the bubble expansion.
     A neck connects the binary and bubble, and is surrounded by an
     oblique shock.  At 1 Myr (bottom), the {\em shape} is similar to
     that at 100,000 years, except that the neck is more elongated,
     with a smaller opening angle.}
\label{fig:slice}
\end{figure}

The dynamical evolution described above is shown in
Fig.~\ref{fig:slice} as a time sequence of density slices through the
center of the simulation box.
Relativistic, axisymmetric simulations of a pulsar wind nebula in
\citet{ber09} created a similar structure with a bow shock connected
to an expanding bubble.  This indicates that, away from the jet or
wind source, the evolution of XRBs and pulsar nebulae are nearly
identical, except for the scale.

\subsection{Scaling relations}

Based on the observed evolution of the trail and bubble, we can
construct a simple analytic model for the inflation of bow shock
nebulae by microquasars (and, by extension, pulsar wind bow shock
nebulae) that can be tested directly against the simulations.  This
will allow us to formulate analytic expressions for observables.

The jets initially inflate a nearly spherical cavity that evolves as a
wind driven bubble \citep{castor:75}.  The bubble expands as
\begin{align} \label{eqn:rb}
  R_{b}(t) = C_1^{1/5} \left(\frac{L}{\rho_0}\right)^{1/5} t^{3/5} 
\end{align}

\noindent
where $L$ is the source luminosity, $\rho_0$ is the ISM density and
$C_1$ is a constant that equals $\frac{25}{14\pi}$ for an adiabatic
index of $\Gamma = 5/3$.  
In terms of the fiducial parameters for our simulations, this is

\begin{align} \label{eqn:radi}
  R_{b}(t) = 13.1 \textrm{~parsec}
  \left(\frac{L_{37}}{\rho_{0H}}\right)^{1/5} t_{5}^{3/5}
\end{align}

\noindent 
where $L_{37} = \frac{L}{10^{37} \textrm{~erg s$^{-1}$} }$, $\rho_{0H}
= \frac{\rho_0}{1.67\times10^{-24} \textrm{~g cm$^{-3}$} }$, and
$t_{5} = \frac{t}{100,000 \textrm{~years} }$.  The expansion velocity
of the bubble is

\begin{align} 
  v_{b}(t) = {\textstyle \frac{3}{5}} C_1^{1/5}
  \left(\frac{L}{\rho_0}\right)^{1/5} t^{-2/5} \label{eqn:vb}
  = 76.7 \textrm{~km s$^{-1}$}
  \left(\frac{L_{37}}{\rho_{0H}}\right)^{1/5} t_{5}^{-2/5}
\end{align}

\noindent
and the pressure inside the bubble is
\begin{align} 
  P_{b}(t) = P_0 \left({\textstyle \frac{5}{4}} M_b^2 - {\textstyle
      \frac{1}{4}}\right) \label{eqn:Pb}
\end{align}

\noindent
where $M_b = \frac{v_b}{\sqrt{\gamma P_0/\rho_0}}$ is the Mach number
of the spherical shock and $P_0$ is the ISM pressure.  Assuming at
least a moderately strong shock, the pressure goes to
\begin{align} 
  P_{b}(t) \approx {\textstyle \frac{5}{4}} P_0 M_b^2 \approx
  {\textstyle \frac{27}{100}} C_1^{2/5} L^{2/5} \rho_0^{3/5}
  t^{-4/5} \label{eqn:Pb_2} \approx 8.25\times10^{-11} \textrm{~erg
    cm$^{-3}$} L_{37}^{2/5} \rho_{0H}^{3/5} t_{5}^{-4/5}
\end{align}

Initially, the expansion of this bubble is faster than the velocity of
the XRB.  However, as the expansion of the bubble slows while the
velocity of the source remains constant, the source eventually breaks
out of the bubble at a time $t_{\rm break}$.  The breakout time scales
with source velocity $v_s$ as
\begin{eqnarray}
  t_{\rm break} & = & C_1^{1/2} \left(\frac{L}{\rho_0}\right)^{1/2} v_s^{-5/2}
  \nonumber \\
  & = & 185,000 \textrm{~years} \left(\frac{L_{37}}{\rho_{0H}}\right)^{1/2} \left(\frac{v_s}{100 \textrm{~km s$^{-1}$}}\right)^{-5/2}
\end{eqnarray}

\noindent
The radius of the bubble at the breakout time therefore scales as
\begin{align}
  R_{break} = C_1^{1/5} \left(\frac{L}{\rho_0}\right)^{1/5}
  t_{\rm break}^{3/5} = C_1^{1/2} \left(\frac{L}{\rho_0}\right)^{1/2}
  v_s^{-3/2}
  = 18.9 \textrm{~parsec} \left(\frac{L_{37}}{\rho_{0H}}\right)^{1/2} \left(\frac{v_s}{100 \textrm{~km s$^{-1}$}}\right)^{-3/2}
\end{align}

After breakout, the source continues to power bubble expansion.  A bow
shock develops in front of the source which bends the jets back in the
opposite direction of the source motion.  A low-density channel
remains which connects the source to the bubble.  Jet material flows
back through this channel, adding energy to the interior of the
expanding bubble.  So long as the volume of this channel remains small
compared to the volume of the bubble, the expansion rate of the bubble
continues to be described by eqn. \ref{eqn:rb}, the self-similar
equation of a continuously powered bubble.

As seen in Fig. \ref{fig:slice}, the resulting shape is an expanding
spherical bubble connected to the source by a thin neck.  The neck
consists of a shock surrounding a narrow cavity filled with jet
material.  At any point $x$ along the neck, the shape is described by
$w_1$, the width of the inner cavity, and $w_2$, the distance from the
mid plane to the outer edge of the shock.  In the frame of the binary,
the cavity has a fixed shape and is in pressure balance with the
surrounding shocked material.  The flow of material can therefore be
described using an adiabatic equation of state, the Bernoulli equation
and mass continuity:
\begin{align}  \label{eqn:adiabatic}
P_x = a \rho_x^{\gamma}
\end{align}
\begin{align}  \label{eqn:bernoulli}
{\textstyle \frac{1}{2}} v_x^2 + \frac{\gamma}{(\gamma-1)} \frac{P_x}{\rho_x} = b
\end{align}
\begin{align} \label{eqn:continuity}
\rho_x v_x w_1^2 = c
\end{align}

\noindent where $v_x$, $\rho_x$ and $P_x$ are the velocity, density
and pressure of material in the cavity at position $x$, $\gamma$, the
adiabatic index, is $5/3$, and $a$, $b$ and $c$ are constants.
In term of $a$, and with $\gamma = 5/3$, $b = a^{3/5} (\rho_0
v_s^2)^{2/5}$ and $c = L \: a^{-3/5} (\rho_0 v_s^2)^{-2/5}$.

Rearranging eqns. \ref{eqn:adiabatic} to \ref{eqn:continuity} we solve
for $w_1$ and find that

\begin{align}
  w_1 = \left( \frac{c a^{3/5}}{\sqrt{2 a^{3/5}}} \right)
  \frac{1}{\left[\left(\rho_0 v_s^2\right)^{2/5} P_x^{6/5} -
      \frac{5}{2} P_x^{8/5}\right]^{1/4} }
\end{align}

\noindent Making the approximation $P_x \ll \rho_0 v_s^2$ (i.e., the
source velocity is supersonic), this reduces to

\begin{align}
  w_1 = 5^{-1/4} 2^{1/2} L^{1/2} a^{3/10} (\rho_0 v_s^2)^{-3/10}
  P_x^{-3/10} = C_2 P_x^{-3/10} \label{eqn:w_1}
\end{align}

\noindent The pressure will be set by the jump conditions for an
oblique shock, with $\theta$ and $\beta$ the angle with respect to the
direction of motion of the inner and outer edges of the shock,
respectively:

\begin{align}
  \tan(\theta) = \frac{M_0^2 \sin^2\beta - 1}{\tan\beta (1 +
    \frac{(\gamma+1)}{2} M_0^2 - M_0^2 \sin^2\beta)} \\
  \frac{P_x}{P_0} = \frac{2 \gamma M_0^2 \sin^2\beta -
    (\gamma-1)}{(\gamma+1)}
\end{align}

\noindent where $M_0 = v_s / \sqrt{\gamma P_0/\rho_0}$ is the Mach
number of the binary relative to the ISM, $\tan\theta = dw_1/dx$ and
$\tan\beta = dw_2/dx$.  Making the approximation that $\theta$ and
$\beta$ are small, this reduces to

\begin{align}
  \theta = \frac{M_0^2 \beta^2 - 1}{\frac{4}{3} \beta M_0^2} \\
P_x = P_0 (\textstyle{\frac{5}{4}} M_0^2 \beta^2 - \textstyle{\frac{1}{4}}) \\
\beta = \sqrt{\frac{4 \frac{(w_1/C_2)^{-10/3}}{P_0} + 1}{5 M_0^2}}
\end{align}

Substituting we find

\begin{align} \label{eqn:dw_1} \theta = \frac{dw_1}{dx} =
  \frac{3(C_2^{10/3} - P_0 w_1^{10/3})}{M_0 P_0 w_1^{10/3}} \left(
    \frac{20 C_2^{10/3}}{P_0 w_1^{10/3}} + 5 \right)^{1/2}
\end{align}

This differential equation can then be integrated numerically to find
$w_1(x)$ and this in turn can be used to find $w_2(x)$.

As the pressure drops along the neck, however, it eventually reaches a
minimum value equal to the pressure in the expanding bubble.  Beyond
this point, the angle of the shock is constant such that the
post-shock pressure is equal to the $P_b$, the pressure in the bubble,
which is set by the expansion rate of the bubble $v_b$.  The incoming
velocity towards the shock will be $v_s - v_b$, because the bubble is
expanding towards the source, giving a minimum shock angle of

\begin{align}\label{eqn:beta}
  \beta = \frac{v_b}{v_s-v_b} = \frac{dw_2}{dx}
\end{align}

\noindent The angle of the inner surface is then approximately

\begin{align}
  \theta = \frac{M_2^2 \beta^2 - 1}{\frac{4}{3} M_2^2 \beta} = \frac{dw_1}{dx}
\end{align}

\noindent where $M_2 = (v_s - v_b) / \sqrt{\gamma P_0/\rho_0}$ is the
Mach number of the shock.  In our simulations, this asymptotic limit
dominates and the shape described by eqn.~\ref{eqn:dw_1} only
describes a small region near the jet source.  Note that the width of
the neck where it reaches the minimum value of $\beta$ is described by
eqn. \ref{eqn:w_1} with $P_x = P_b$, giving $w_1(P_b) = C_2
P_b^{-3/10}$.  This equation depends on the normalization constant $a$
in the adiabatic equation (eqn. \ref{eqn:adiabatic}), which in
practice depends on the amount of mixing between jet and ISM material.
The width at this point also depends on the radius of curvature of the
bow shock, which is not taken into account in this analytic model
Therefore, we use $a$ as a free parameter to get the proper fit for
the width of the neck where the minimum value of $\beta$ is reached.

\begin{figure}[!htbp]
   \centering
   \includegraphics[width=0.6\textwidth]{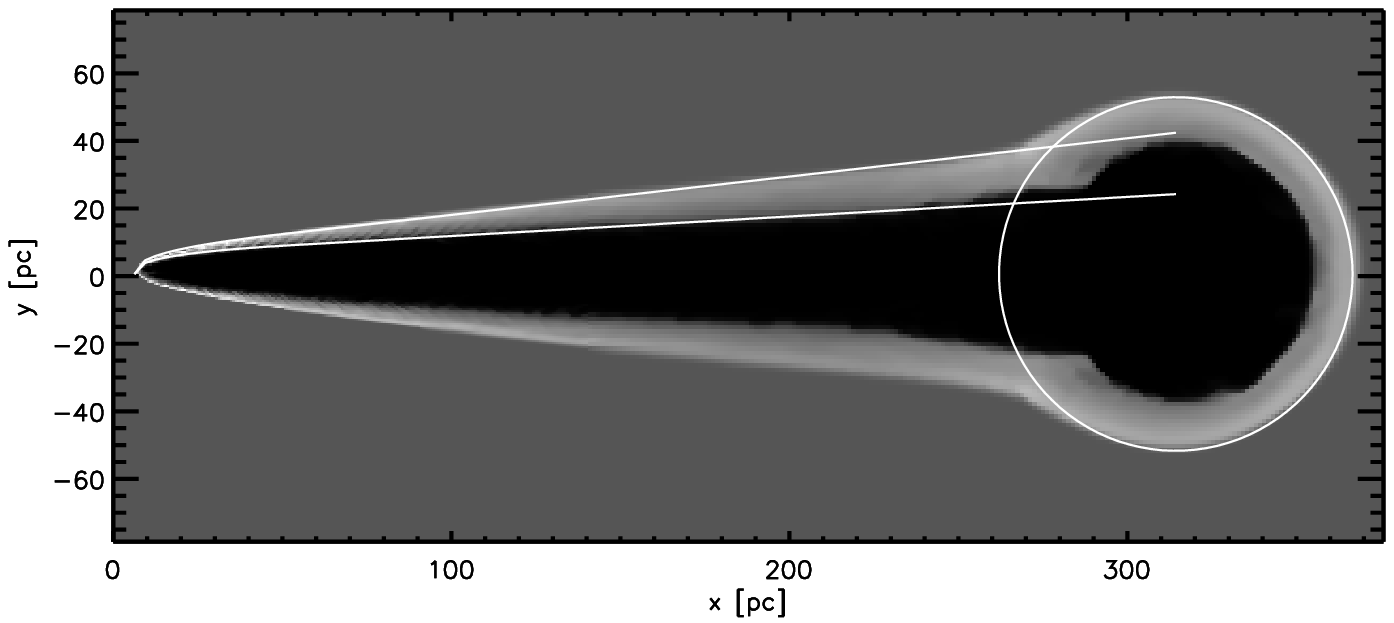}\\
   \includegraphics[width=0.3\textwidth]{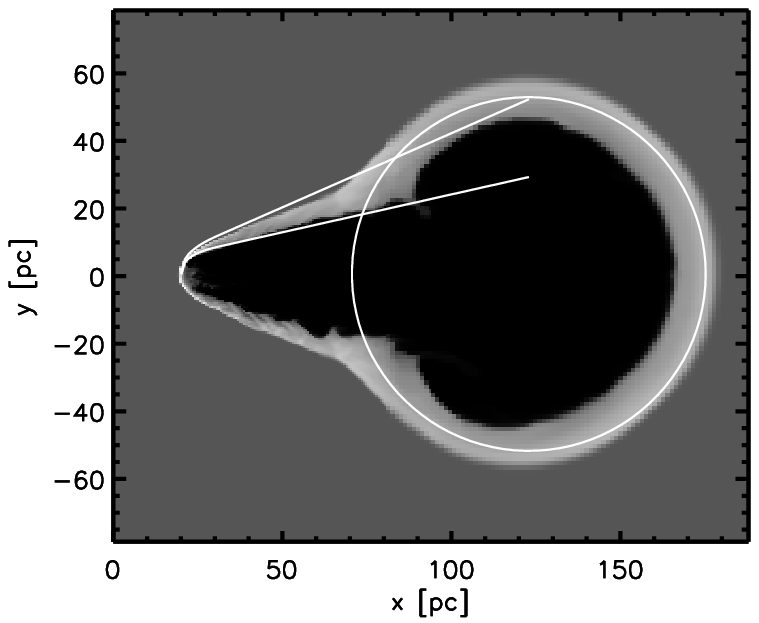}
   \includegraphics[width=0.3\textwidth]{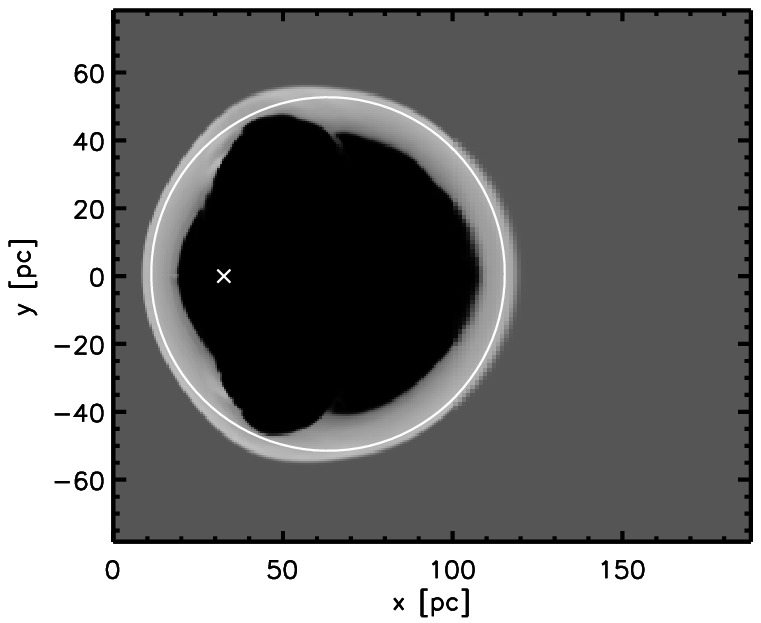}
   \caption{Density slice of 300 ~km s$^{-1}$ (upper), 100 ~km s$^{-1}$ (lower left) and 30 ~km s$^{-1}$ (lower right)
     simulations at 1 Myr.  The white curves trace the analytic
     solution for $w_1$ and $w_2$, the neck shape, and the circle is
     the size of the bubble from eqn. \ref{eqn:rb}. Values of $a$ used
     to fit the bow shock width are $10^{31}$ for the $300$ ~km s$^{-1}$
     simulations and $10^{33}$ for the $100$ ~km s$^{-1}$ simulation. 
     In the 30 ~km s$^{-1}$ simulation the XRB (marked with an $\times$) is 
     still inside the bubble, but significantly offset from the center.}
\label{fig:100and300}
\end{figure}

Figure \ref{fig:100and300} shows density slices of the 300 ~km s$^{-1}$, 100
~km s$^{-1}$ and 30 ~km s$^{-1}$ simulations (models hv, mv and lv) at 1 Myr.  The
white lines are results of our analytic model for the bubble size,
$w_1(x)$ and $w_2(x)$. In the first two cases, the predicted shape of
the inner and outer edge of the neck are a good fit to the
simulations.  In the 30 ~km s$^{-1}$ case, the XRB is still inside the
spherical bubble at 1 Myr.

The outer shock angle $\beta$ decreases with time as the pressure in
the bubble and the expansion velocity drop.  The width of the neck
where it meets the bubble is approximately

\begin{align}
  w_{2,meet} = \beta (v_s t - R_b) + w_1(P_b) \\
  w_{2,meet} = \frac{v_b}{v_s - v_b} \left(v_s t -
    \textstyle{\frac{5}{3}} v_b t\right) + C_2 \left(
    \textstyle{\frac{5}{4}} P_0 M_b^2 \right)^{-3/10}
\end{align}

Asymptotically, this width goes to $w_{2,meet} = \beta v_s t = v_b t =
\frac{3}{5} R_b$.  Therefore, the ratio of the bubble radius to the
neck width approaches a constant ratio of $3/5$, and the width of the
neck is always smaller than the radius of the bubble.

The volume of the neck scales asymptotically as

\begin{align}
  V_{\rm neck} = \textstyle{\frac{1}{3}} \pi v_s t
  \left(\textstyle{\frac{3}{5}} R_b \right)^2 =
  \textstyle{\frac{1}{5}} \pi C_1^{2/5} v_s \left( \frac{L}{\rho_0}
  \right)^{2/5} t^{11/5} = 3.23\times10^{58} \textrm{~cm$^{3}$}
  \left(\frac{v_s}{100 \textrm{~km s$^{-1}$}}\right) \left(
    \frac{L_{37}}{\rho_{0H}} \right)^{2/5} t_{5}^{11/5}
\end{align}

\noindent while the volume of the bubble scales as

\begin{align}
  V_{bub} = \textstyle{\frac{4}{3}} \pi R_b^3 =
  \textstyle{\frac{4}{3}} \pi C_1^{3/5} \left( \frac{L}{\rho_0}
  \right)^{3/5} t^{9/5} = 2.75\times10^{59} \textrm{~cm$^{3}$} \left(
    \frac{L_{37}}{\rho_{0H}} \right)^{3/5} t_{5}^{9/5}
\end{align}

\noindent the ratio of the volumes is therefore

\begin{align}
  \frac{V_{\rm neck}}{V_{bub}} = \frac{9}{100} \frac{v_s t}{R_b} =
  \frac{3}{20} C_1^{-1/5} v_s \left( \frac{L}{\rho_0} \right)^{-1/5}
  t^{2/5} = 0.117 \left(\frac{v_s}{100 \textrm{~km s$^{-1}$}}\right)
  \left( \frac{L_{37}}{\rho_{0H}} \right)^{-1/5} t_{5}^{2/5}
\end{align}

The volume of the neck and bubble will eventually become equal when
the length of the neck ($v_s t$) is about $11.1$ times the bubble
radius.  The width of the shock where it meets the bubble is about
$w_{2,meet} = \frac{3}{5} R_b$, so the angle when the volumes are
equal is $\beta = 27/500 = 0.054$ However, the angle of the shock
around the neck cannot be less that $\beta \simeq 1/M_0$, so the
volumes become equal before the shock becomes weak only if $M_0 \geq
18.5$.  Our assumption that the volume of the neck is small will hold
until bubble expansion starts to become marginally sonic, unless the
source has a very high Mach number relative to its surroundings.

While our simulations have not run long enough 
to probe the sub-sonic regime
of bubble expansion, it is worth speculating about the late state
evolution of trails and bubbles.  Given that the bounding pressure of
the bubble will be dominated by the internal pressure of the ISM, the
expansion velocity of the bubble will drop below the self-similar
value for an energy driven bubble (with $R \propto t^{1/3}$).  In
addition, the ISM will no longer be strongly compressed into a narrow
shell.  Since the neck and bow shock will maintain their stationary
shape, one should expect that the late state evolution of a
bow-shock/trail nebula will eventually lose the terminating bubble and
the trail pressure will eventually approach the ISM pressure,
consistent with the late state structure of the trail proposed in
\citet{hei08}.

\subsection{Observable properties of microquasar bow shock nebulae
  and trails}

\subsubsection{H$_\alpha$ emission}
For XRBs moving supersonically through cold or warm ISM, the
compressed, hot gas in the bow shock will produce collisionally
excited line emission, most importantly H$_\alpha$ (as well as the
classic spectrum of nebular lines like [OIII]).  In fact,
\citet{wie09} detected the nebula of SAX J1712.6-3739 in H$_\alpha$.
Our models support the jet-ISM interaction scenario \citep{hei08} and
the numerical simulations show an apparent quantitative agreement with
the observed results for H$_\alpha$ emission.

\begin{figure}[!htbp] 
  \begin{center}$
    \begin{array}{cc}
    \includegraphics[width=0.5\textwidth]{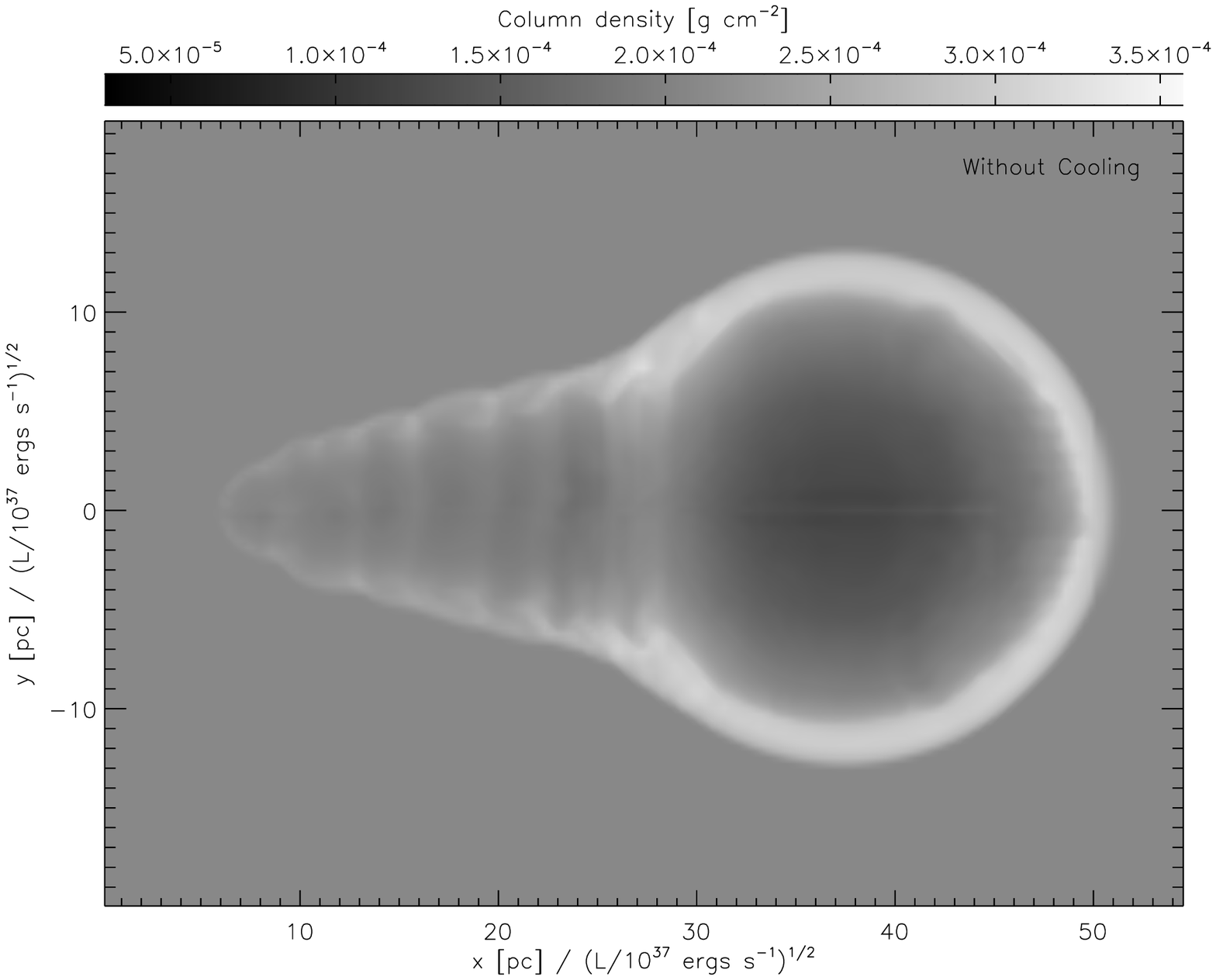} &
    \includegraphics[width=0.5\textwidth]{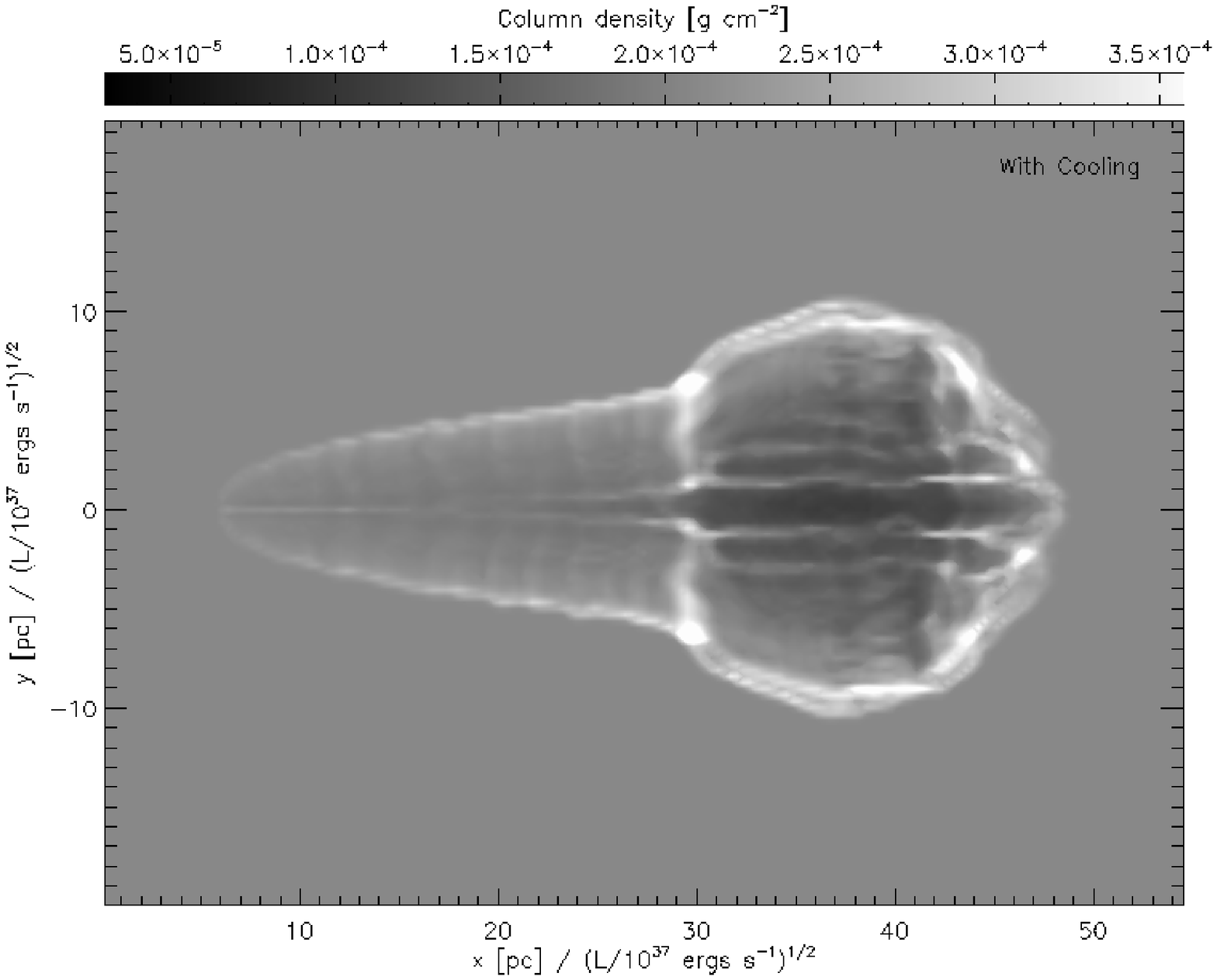} \\ 
    \includegraphics[width=0.5\textwidth]{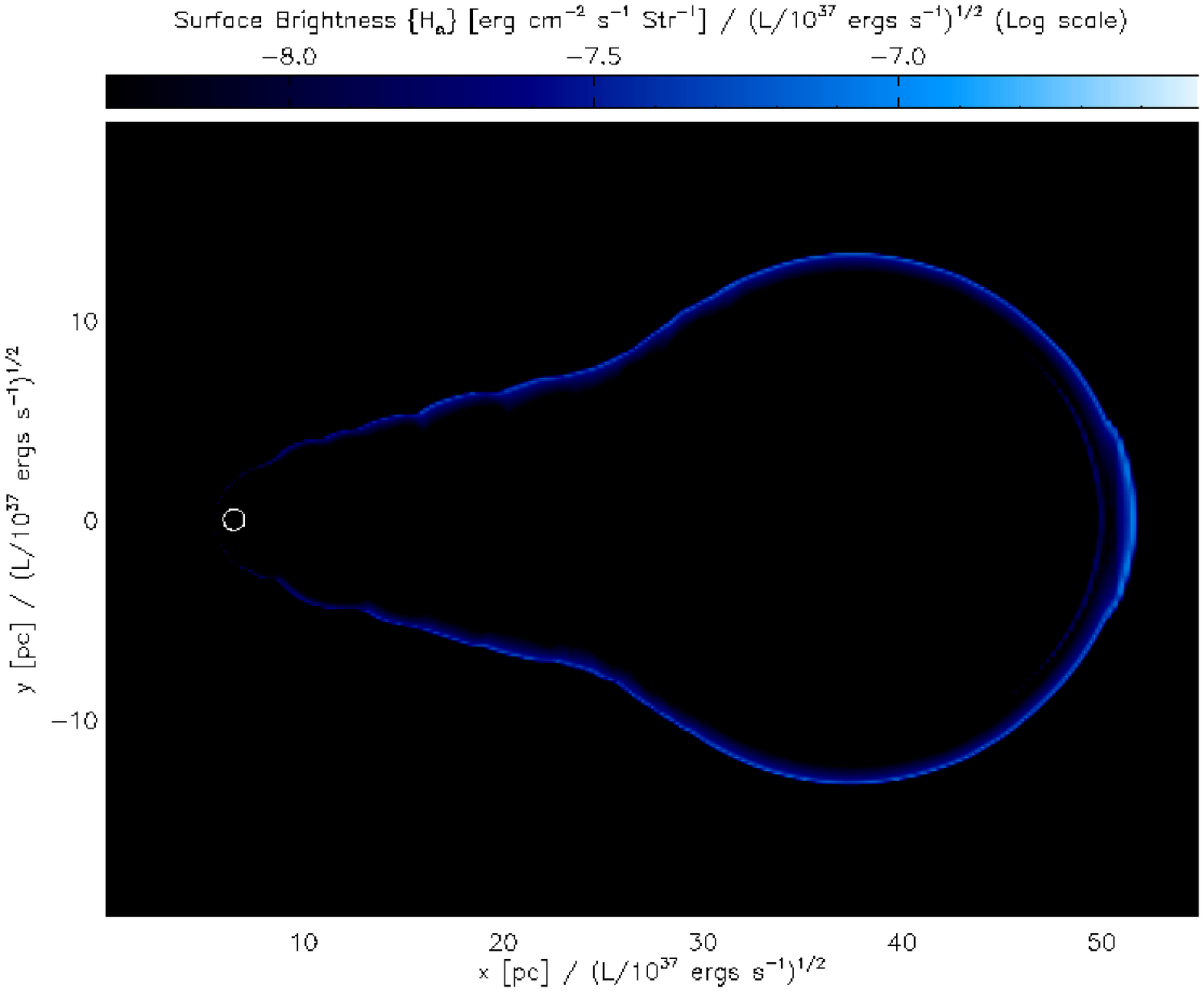} &
    \includegraphics[width=0.5\textwidth]{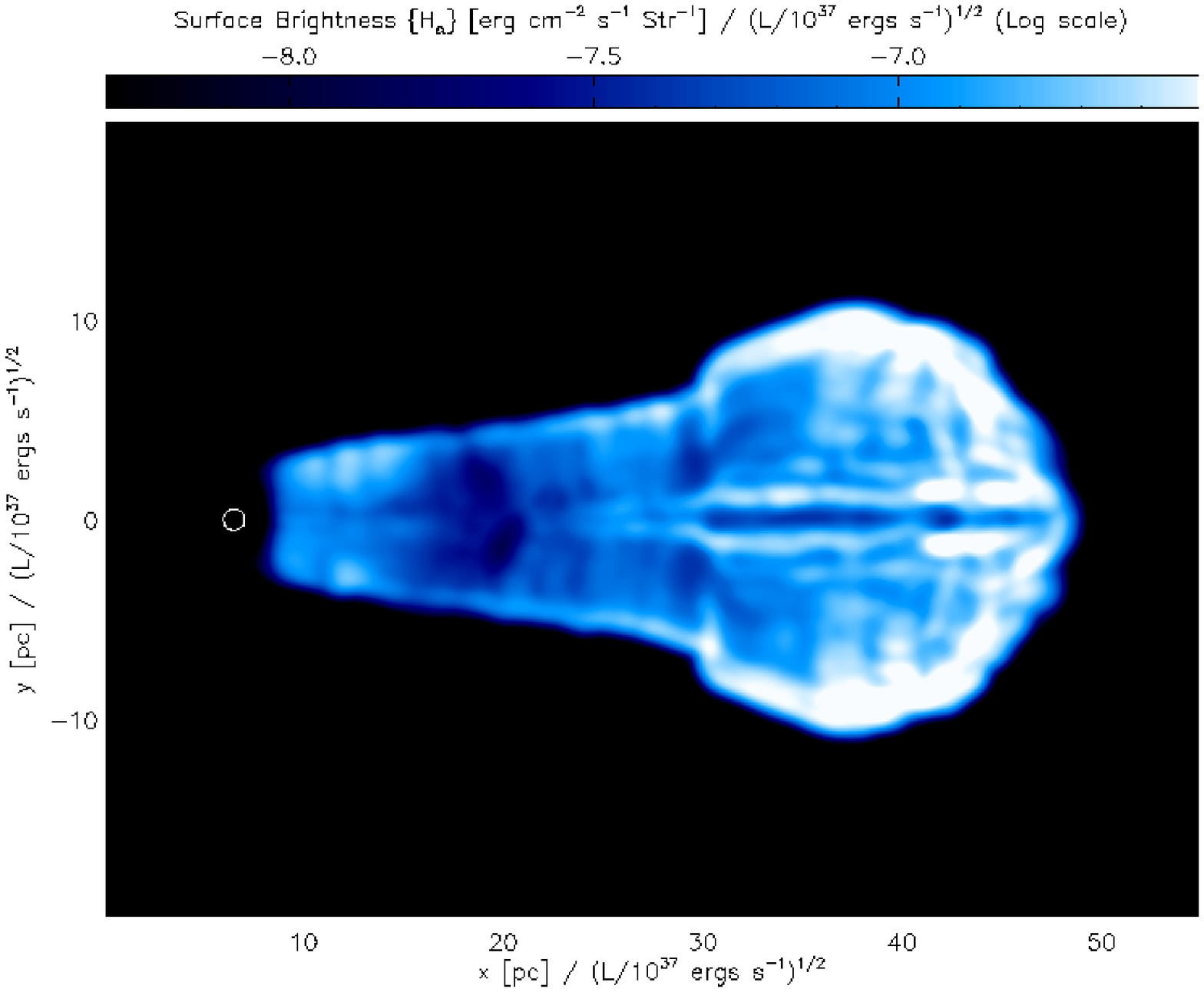}  
 \end{array}$ 
\end{center} 
\caption{The density contour map for the model without the radiative
  cooling (upper left) and with the radiative cooling (upper right) at
  100,000 yr (v$_{xrbs}=300 \textrm{~km s}^{-1}$). The bottom panels shows
  H$_\alpha$ emission for each case (log scale). The predicted surface
  brightness can be scaled to different XRB jet luminosities by
  reducing the surface brightness, size scale and simulation time by a
  factor of L$_{37}^{-1/2}$.}
\label{fig:halpha} 
\end{figure}

To calculate the H$_\alpha$ emission, we first determine the
ionization balance in the shocked gas, using the MAPPINGS III code
\citep{sut93}; the code uses a time-dependent algorithm for accurate
equilibrium balance calculation.  The typical temperature of the
shocked shell is estimated to be around $10^{5} - 10^{6}$ K in the
adiabatic case, hence the gas is inferred to be nearly fully ionized
at the shocked region.  Figure \ref{fig:halpha} (bottom left panel)
shows the surface brightness in H$_\alpha$ for the model hv.  Note
that all simulations can be scaled with XRB jet luminosity by reducing
the predicted surface brightness, nebula size, and simulation time by
a factor of $\left(L_{37}\right)^{-1/2}$.

The bubble is relatively bright in H$_\alpha$.  The shock around the
neck is also visible in H$_\alpha$ in fig. \ref{fig:halpha}.  A
similar feature has been identified with SAX J1712.6-3739
\citep{wie09}. We discuss the morphological similarity between that
source and our simulations in \S 4.1.

Radiative cooling has little influence on the dynamical evolution of
the XRBs.  The upper panels of fig.~\ref{fig:halpha} reveal that
overall morphologies for two models (with and without radiative
cooling) are broadly similar.  Radiative cooling leads to a
significantly thinner shell of shocked material around the bubble and
neck, giving the appearance of a slightly narrower neck.  Because of
the cooling-induced contraction of the gas, the shell also appears to
develop some irregularities in shape, though the overall shape of the
hot (radio) plasma inside the trail and bubble occupies essentially
the same volume.

However, the radiative cooling has a pivotal role for H$_\alpha$,
because of the strong temperature sensitivity of the ionization
balance and thus the recombination line emission.  This is because the
cooling time scale is comparable to the dynamical time scale of XRBs,
leading to significantly lower temperatures in the shell than in a
purely adiabatic simulations.  

The bow shock immediately surrounding the XRB is very dim because the
temperature is too high for strong H$_\alpha$ emission.  As cooling
becomes important down stream, the temperature at the shock quickly
drops to the range of a few 10,000 K.  As a result, the emission
becomes stronger by about a factor of 50.

Because H$_{\alpha}$ emission depends non-trivially on temperature, it
is not straight forward to express brightness predictions
semi-analytically.  Figure \ref{fig:halpha} is therefore specific to
our simulation using fiducial parameters only.

\subsubsection{Radio emission}
Given that the trail itself should be filled by magnetized,
relativistic plasma released by the jets, they will emit synchrotron
radiation.  As the plasma travels along the jet, particles will cool
both adiabatically and due to synchrotron losses.  As is well known
from studies of AGN jets and radio lobes, this introduces a cutoff to
the electron energy distribution and to the synchrotron spectrum.

Following, e.g., \cite{hei97}, the cutoff frequency, as a function
of travel time along the trail, will be given by

\begin{equation}
  \gamma_{max} \simeq \frac{\left( \frac{P}{P_{0}} \right)^{1/4}}
  {\int{\frac{4}{3}\sigma_{T}U_{B} \left( \frac{P}{P_{0}} \right)^{1/4} dt}}.
\end{equation}

\noindent with an associated cutoff frequency of

\begin{equation}
  \nu_{max} \approx \frac{9 e B \gamma_{max}^{2}}{4 \pi m_{e} c^{2}},
\end{equation}

\begin{figure}[!htbp]
\plottwo{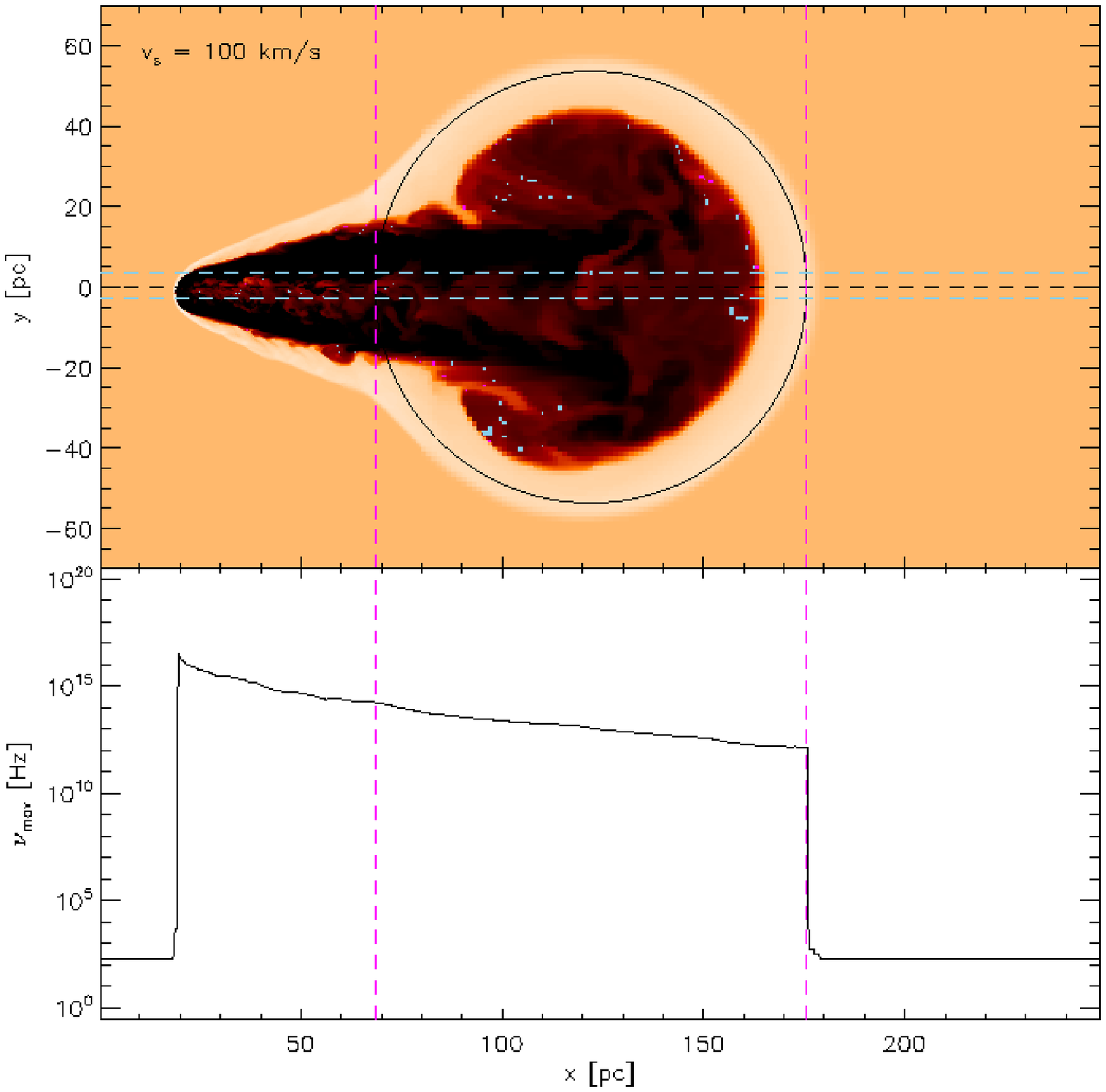}{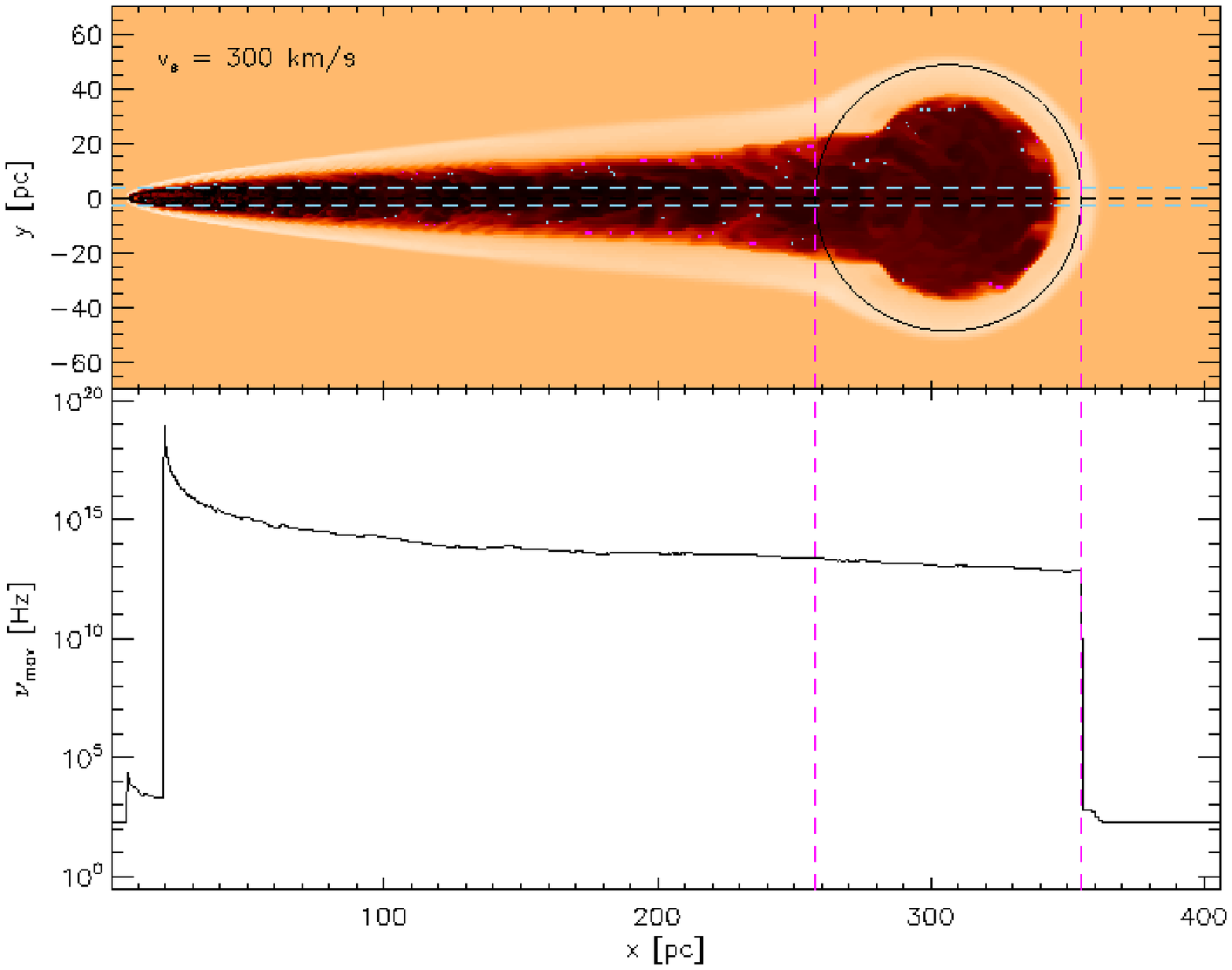}
\caption{Synchrotron cooling frequency for X-ray binaries at 100 ~km s$^{-1}$
  (left panel) and 300 ~km s$^{-1}$ (right panel) at 1 Myr. Top images
  indicate the density map with logarithmic scale.  The vertical
  magenta lines indicate edges of the bubble area (marked with the
  solid circle).  The cutoff frequency is calculated by integrating
  along the two blue lines and assuming a steady
  state. \label{fig:sync}}
\end{figure}

Given that the flow through the trail assumes a quasi-steady state
behind the XRB, we can simply integrate this equation through single
frames to lowest order to derive the cooling frequency as a function
of position along the trail.  The result is shown in
fig. \ref{fig:sync}.  For our fiducial parameters, the cooling
frequency is estimated to fall within $10^{13} - 10^{15}$~Hz, well
above the radio band.

This implies that synchrotron emission should be a good tracer of
these trails at all frequencies, not just at low frequencies.  This is
in contrast to the estimates in \cite{hei08}, who suggested that
cooling could be important along the trail.  The reason for this
difference is the significant backflow velocity along the trail, which
was left as a free parameter in \cite{hei08}, and which allows radio
emitting particles to traverse the trail without significant losses.

\begin{figure}[!htbp] 
\begin{center}$ 
  \begin{array}{c}
    \includegraphics[width=0.5\textwidth]{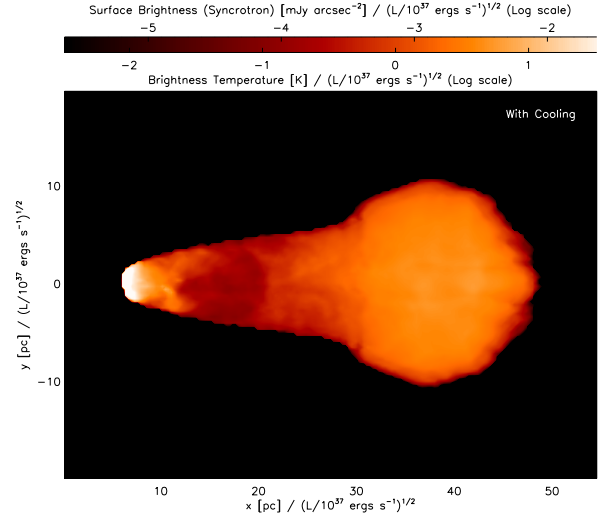} 
 \end{array}$ 
\end{center}
\caption{Surface brightness of radio synchrotron emission for the
     frequency of 1 GHz at 100,000yr ($v_{\rm XRB}=300$ ~km s$^{-1}$). 
   }
\label{fig:radio} 
\end{figure}

Figure \ref{fig:radio} shows synchrotron surface brightness for the
$300$ ~km s$^{-1}$ simulation at 100,000 years.  Surface brightness for
synchrotron emission is strongest at the front-edge where the jets are
bent by ram pressure.  The terminal radio bubble is also bright and
emits the bulk of the total radio luminosity.

In fig. \ref{fig:radio}, the typical brightness temperature inside the
bubble is estimated to be of order a few K in model hvc at 0.1 Myr at
a frequency of 1 GHz.  Since the synchrotron emission depends only on
the pressure, the typical values of brightness temperature in other
models ($v_{\rm XRB}$=30, 100~km s$^{-1}$) are essentially the same.

If the intensity of synchrotron emission is high enough, synchrotron self-absorption will occur and the emission will
drop out at low frequencies, proportional to $\nu^{5/2}$ regardless of the electron power index. 
However, in our particular simulation, the optical depth at 1 GHz is quite transparent (about $\tau = 10^{-2}$)
so this effect would be negligible.

The bright temperature can be extrapolated into a few $10^{-3}$ K in
submm wave bands that might be detectable by a new generation detectors
such as SCUBA-2 or ALMA. In the case of the XRBs in our galaxy, assuming 
that the distance would be a few kpc, the angular size of the bubble is 
large enough to be resolved by these detectors.

Since the pressure inside the bubble decreases with time, the surface
brightness will also decrease with time and synchrotron radiation will
be easier to observe at an earlier XRB age if the bubble is resolved.

\begin{figure}[!htbp] 
\begin{center}$ 
\begin{array}{c} 
    \includegraphics[width=0.5\textwidth]{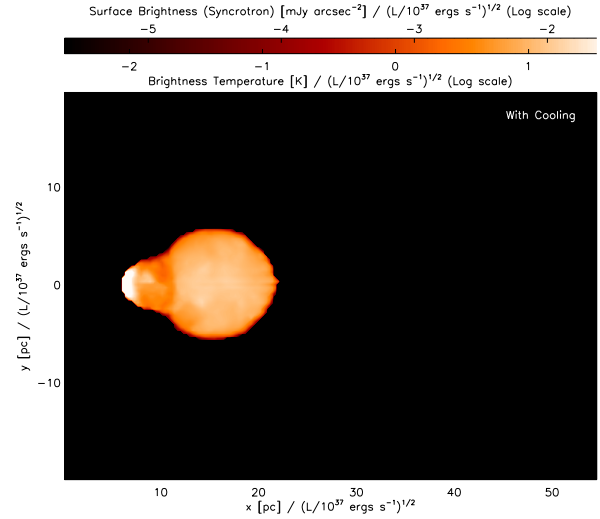} 
 \end{array}$ 
\end{center} 
\caption{Surface brightness of radio synchrotron emission for the
     frequency of 1 GHz at 30,000yr ($v_{\rm XRB}=300$~km s$^{-1}$). 
   }
\label{fig:radio_young} 
\end{figure}

The synchrotron emission from the bubble can easily be predicted
analytically.  For a power-law distribution of electrons,
$N(\gamma)d\gamma = A \gamma^{-p} d\gamma$: if we assume the power-law
index (p) is 2.5, the total emissivity per unit volume per unit
frequency can be calculated as,

\begin{equation}\label{eq:em_sync}
   \epsilon_{sync} = C_{3} P^{15/8},
\end{equation}

\noindent where P is the pressure in the bubble and $C_{3}$ is
approximately 1.44$\times$10$^{-17}$ and depends on the ratio of
electron to magnetic pressure.  Throughout this discussion we will
assume that the plasma is in equipartition (with equal energy density
in electrons and magnetic field), making the estimated fluxes {\em
  upper limits}.

The pressure of the hot bubble can be estimated from the shock jump
conditions. If we set the adiabatic index to 5/3, the pressure jump
condition in the shell of the bubble are given by eqn. \ref{eqn:Pb_2}.

\begin{figure}[!htbp]
   \centering
   \includegraphics[width=0.7\textwidth]{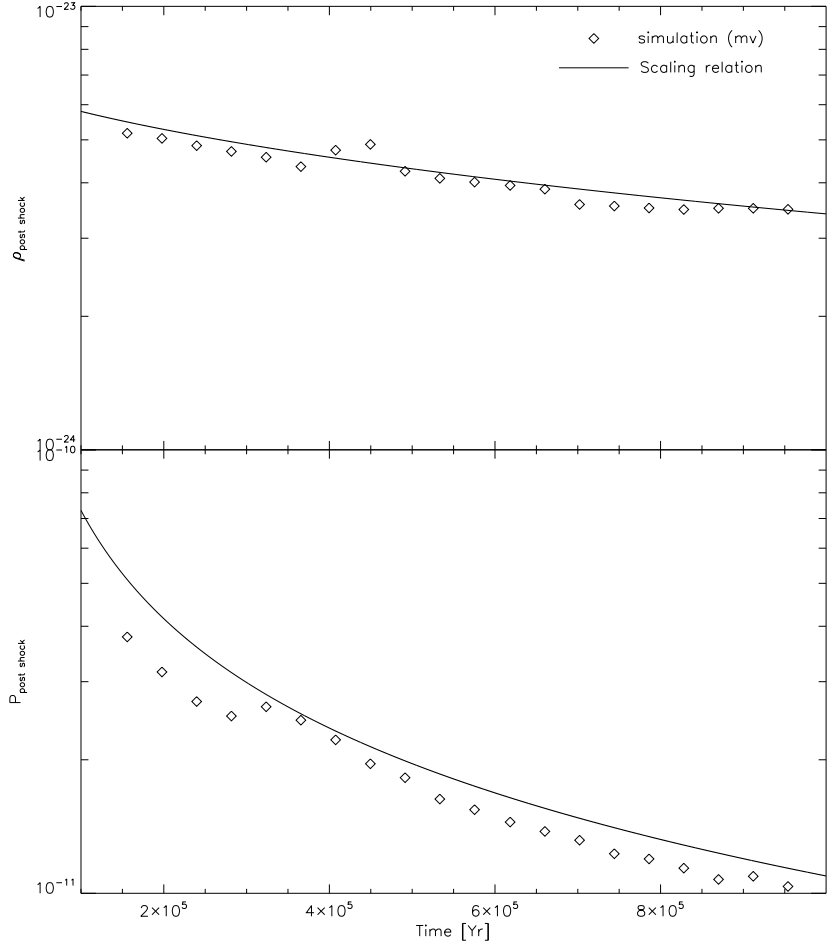}
   \caption{The solid line indicates that post shock pressure and
     density derived from scaling relation and diamond marks the
     results from simulation.}
   \label{fig:evol_pd}
\end{figure}

Figure \ref{fig:evol_pd} shows that the inferred post-shock pressure
is consistent with the simulation results, although the data from the
simulation have a slightly smaller values. The small discrepancy is
negligible in estimating the surface brightness.

The surface brightness can be derived from the emissivity
(eqn.~\ref{eq:em_sync}) with post-shock pressure and the path length
of the line of sight in bubble. Hence the maximum value of path length
is R$_{bub}$(t), and the surface brightness becomes

\begin{eqnarray}
  I_{sync} &\approx& \frac{C_{3}} {4 \pi} \left( \frac{27}{100}
    C_{1}^{2/5}L^{2/5}\rho_{0}^{3/5} t^{-4/5} \right)^{15/8} 
  C_{1}^{1/5}L^{1/5}\rho_{0}^{-1/5}t^{3/5}  \nonumber \\
  &\approx& 1.37\times10^{-3} \textrm{mJy arcsec}^{-2} \times 
  \left( L_{37} \right)^{19/20} \left( \rho_{0H} \right)^{37/40}
  \left( \frac{t}{1 \textrm{Myr}} \right)^{-9/10}
\end{eqnarray}

\begin{figure}[!htbp]
   \centering
   \includegraphics[width=0.7\textwidth]{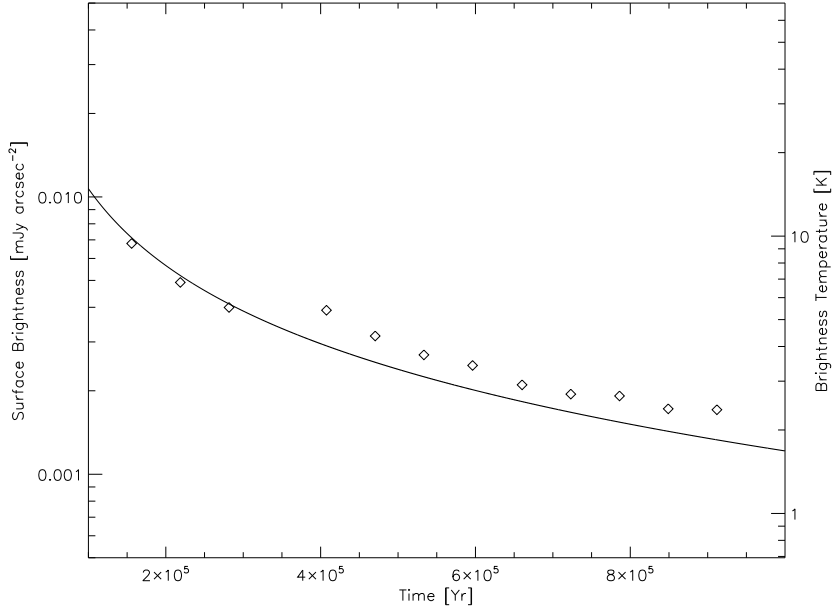}
   \caption{The solid line indicates the theoretically expected
     surface brightness in synchrotron emission and diamond indicates
     the results from simulation.}
\label{fig:surfb_sync}
\end{figure}

\noindent This theoretical calculation is in good agreement with
numerical data shown in fig.~\ref{fig:surfb_sync}.  The relation
confirms that the synchrotron emission decreases over time, implying
that younger XRBs are more easily detected in the radio.

XRB bubbles in other galaxies will be difficult to resolve with radio
observations, but may be detectable as unresolved sources, at
locations {\em different} from the X-ray point source.  If we set the
distance to the XRBs to be D, then we can estimate the unresolved flux
as,

\begin{eqnarray}
  \textrm{Flux}_{\textrm{sync}} &\approx& \frac{C_{3}}{16 \pi} \left(
    \frac{27}{100} C_{1}^{2/5}L^{2/5}\rho_{0}^{3/5} t^{-4/5} \right) 
  ^{15/8} C_{1}^{3/5}L^{3/5}\rho_{0}^{-3/5}t^{9/5} D^{-2} \nonumber \\
  &\approx& 0.04 ~\textrm{mJy} \times \left( L_{37} \right)^{27/20} 
  \left( \rho_{0H} \right)^{21/40} \left( \frac{t}{1 \textrm{Myr}} \right)^{3/10}
  \left( \frac{D}{1 \textrm{Mpc}} \right)^{-2}
\end{eqnarray}

\noindent This flux includes only the bubble area, and neglects the
synchrotron emission from near the jets.  Because we do not resolve
the base of the inner regions of the jet in our simulations, and
because microquasars are generally highly variable, we cannot make any
quantitative statements about the relative flux from the inner jet and
the bubble. Note that the total flux from the trail and bubble {\em
  increases} with time, indicating that older XRBs will be easier to
detect than young ones in unresolved observations.

The state-of-the-art radio observatories are well suited for this type of
objects; the eVLA has been upgraded to have the excellent sensitivity, frequency coverage, and imaging
capability that allows for good quality radio spectra resolved over the source, and the WSRT APERTIF that
has very large field of view at 1.4 GHz, allowing for deep continuum surveys capable of detecting trailed nebula.  

\subsubsection{Free-free emission}

\begin{figure}[!htbp] 
\begin{center}$ 
\begin{array}{c} 
  \includegraphics[width=0.5\textwidth]{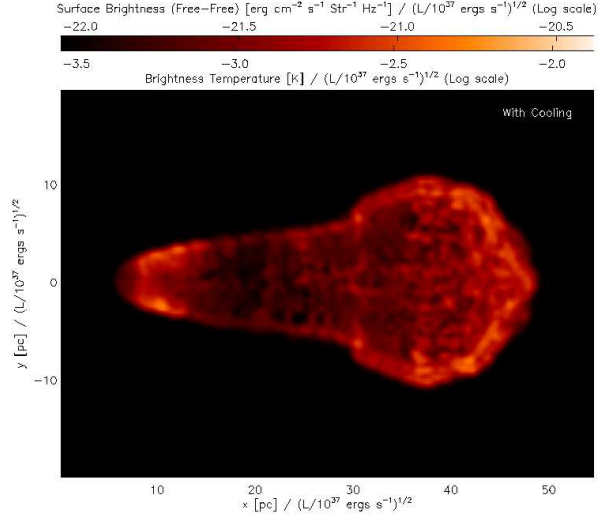} 
 \end{array}$ 
\end{center} 
\caption{Surface brightness of free-free emission for the frequency
     of 1GHz at 100,000 yr (v$_{xrbs}=300$ ~km s$^{-1}$).
}   
\label{fig:free} 
\end{figure}

For the expected temperature range in the bow shock, we can expect the
gas to be mostly or completely ionized, which will give rise to the
emission of Bremsstrahlung.  Figure \ref{fig:free} shows a simulated
radio image in free-free emission for the case of $v_{\rm XRB}=$300 ~km s$^{-1}$.
The emission is edge--brightened, given that it comes exclusively from
the compressed ISM shell.

The temperature at the shell of the bubble can be calculated by the
post-shock density and pressure as
\begin{eqnarray}\label{eq:free_T}
  T_{bub} = \frac{\mu}{k} \frac{P_{1}}{\rho_{1}} \nonumber \approx
  \frac{27}{400} C_{1}^{2/5} \frac{\mu}{k} \rho_{0}^{-2/5} L^{2/5}
  t^{-4/5}, 
\end{eqnarray}
\noindent where $\mu$ is mean molecular mass.  If we assume fully
ionized gas inside the shell, the free-free surface brightness through
the shock can be approximately written as
\begin{eqnarray}
  I_{ff} &\approx& 3.91\times 10^{-20} ~\textrm{ergs
    s}^{-1}\textrm{cm}^{-2}\textrm{Hz}^{-1}\textrm{str}^{-1}  
  \times \left( \rho_{0H} \right)^{2} \left( \frac{t}{1 \textrm{Myr}}
  \right)
\end{eqnarray}
where we assume a strong shock with $\rho_{shock} \approx 4\rho_{0}$
for simplicity.

Free-free emission is dimmer than the synchrotron emission in the
bubble, and is expected to be undetectable or marginally detectable at
$\nu$=1Ghz.  

For the shock temperatures in our simulations ($10^4 - 10^6$ K),
optical emission is dominated by cooling lines rather than free-free
continuum emission.  However, for a fast source the temperature in the
bow shock may initially exceed $10^6$ K, in which case free-free
emission would be the dominant coolant.

\subsubsection{X-ray}

\begin{figure}[!htbp]
   \centering
   \includegraphics[width=0.6\textwidth]{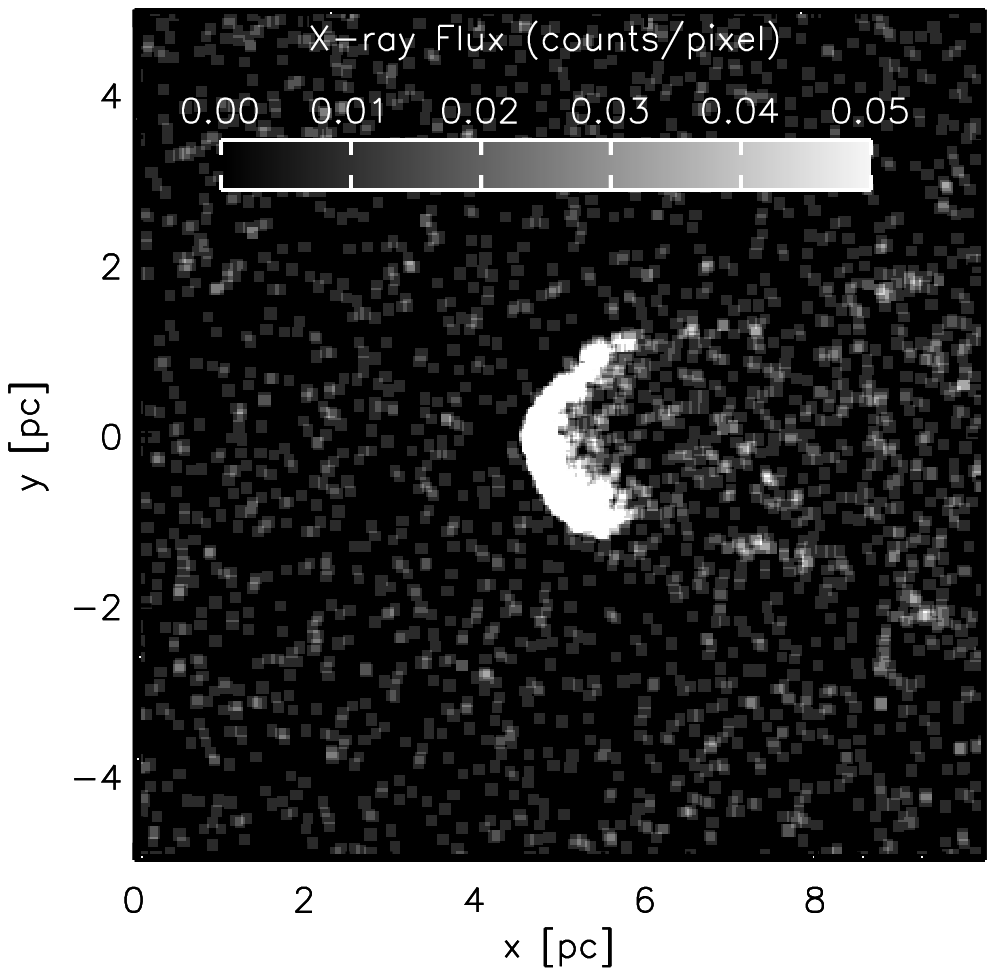}
   \includegraphics[width=0.6\textwidth]{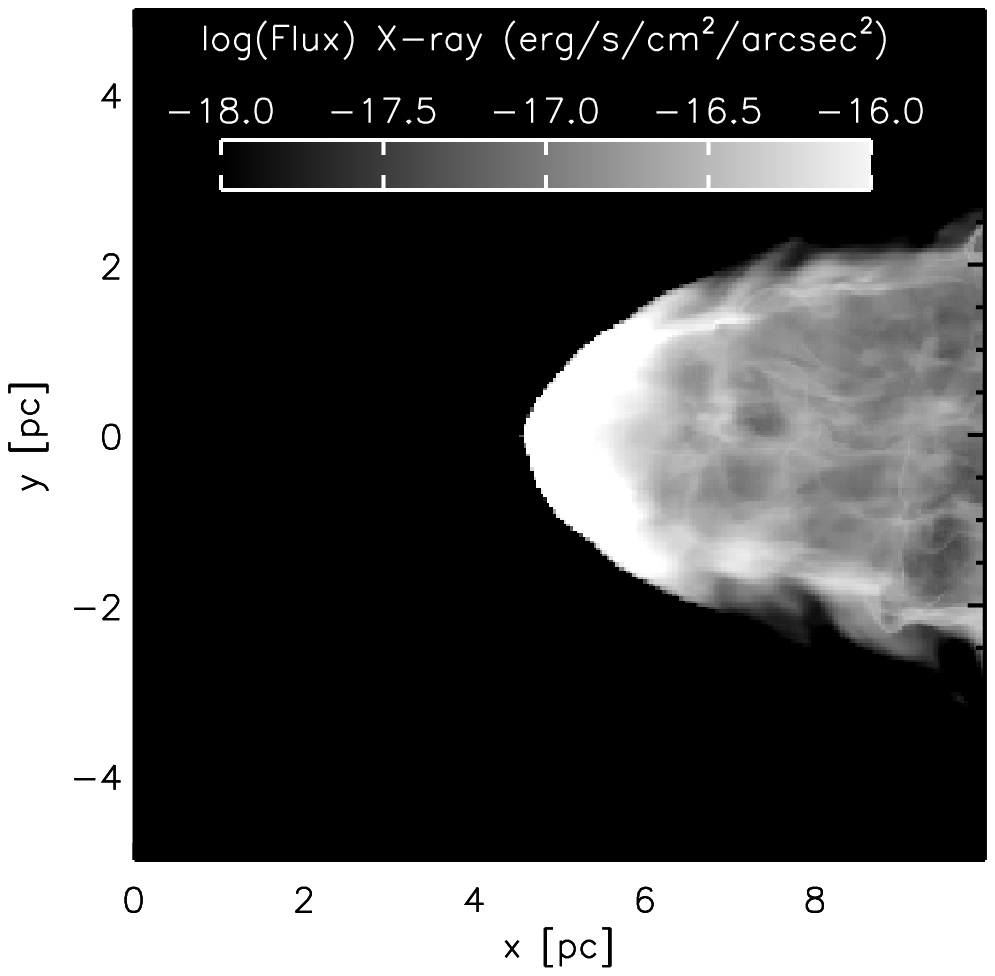}
   \caption{Simulated Chandra X-ray image (top) and surface brightness
     (bottom) of the bow shock for the 300 ~km s$^{-1}$ simulation at
     1~Myr. 100~ks exposure at 5~kpc and flux is integrated from 0.3
     to 3 keV. The maximum flux in the bow shock is about 1
     count/pixel.}
\label{fig:xray_chandra_300_hires}
\end{figure}

For large enough space velocities, the temperatures at the stagnation
point of the bow shock can be sufficiently high to produce X-rays.
Fig.~\ref{fig:xray_chandra_300_hires} shows a simulated Chandra X-ray
observation produced using the XIM program \citep{heinz09} for the 300
~km s$^{-1}$ simulation at a distance of 5~kpc with a 100~ks exposure at
$10^6$ years.

For the fiducial parameters, there is a significant flux only from the
bow shock, which has a temperature of about $2\times10^{6}\,{\rm K}$.
However, it may be difficult to distinguish this flux from the
emission from the XRB.  There is a small amount of X-ray emission from
the neck, but it is only marginally distinguishable from the
background, and fades as you go farther down the neck.

\begin{figure}[!htbp]
  \centering
  \includegraphics[width=0.6\textwidth]{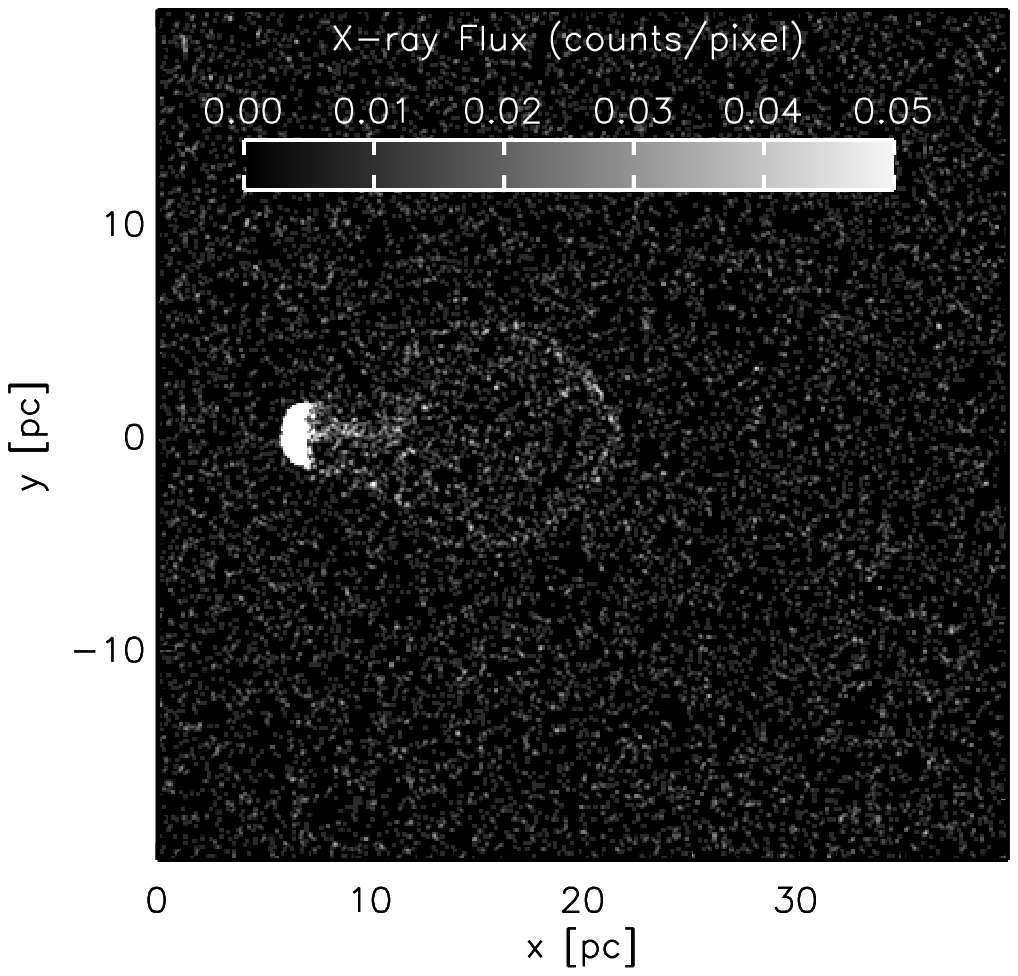}
  \includegraphics[width=0.6\textwidth]{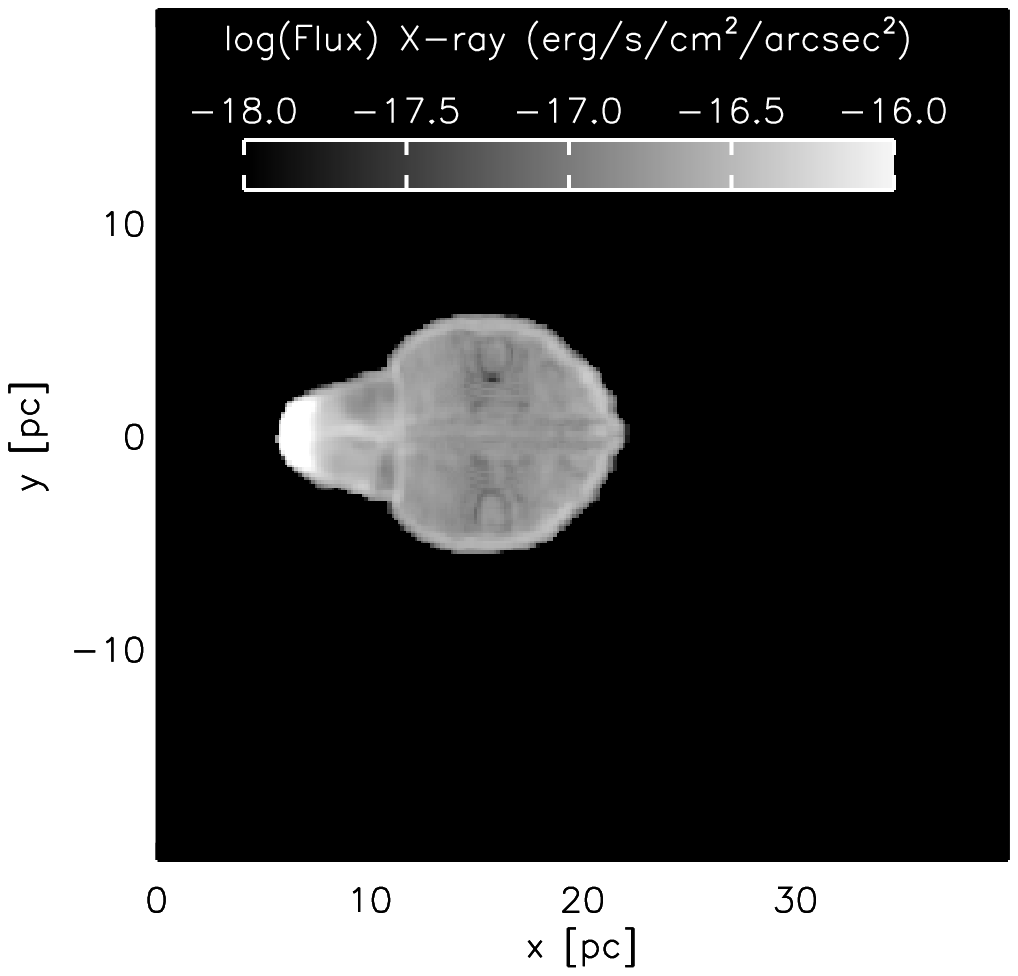}
  \caption{Simulated Chandra X-ray image (top) and surface brightness
    (bottom) of the bubble for the 300 ~km s$^{-1}$ simulation (hvc) at 30,000
    years. 100~ks exposure at 8~kpc and flux is integrated from 0.3 to
    3 keV.}
  \label{fig:xray_chandra_300_lowres}
\end{figure}

At early times, the expanding bubble produces a detectable X-ray flux.
Figure \ref{fig:xray_chandra_300_lowres} shows a simulated Chandra
X-ray observation for the 300 ~km s$^{-1}$ simulations (hvc) at a distance of
8 kpc with a 100~ks exposure at 30,000 years.  Although it has a low
surface brightness, the bubble and neck can be seen, in addition to
the bright bow shock, in a smoothed image.

A shock will only be detectable in X-rays if its temperature is above
about $5\times10^5$ K, which corresponds to a shock velocity of about
100 ~km s$^{-1}$ for the parameters in our simulations.  Assuming a strong
shock, the shock temperature is

\begin{align}
  T_s \sim 6.79\times10^3 K \times M_s^2 \times
  \left(\frac{P_0}{3\times10^{-12}} \right) \left(
    \frac{\rho_0}{1.67\times10^{-24}} \right)
\end{align}

The bubble expansion falls below about 100 ~km s$^{-1}$ at about 36,000 years.
After this, the bow shock is still visible, if the XRB is moving fast
enough, but the bubble and neck are unlikely to be detected.  The bow
shock is always visible, but only for XRBs moving faster than about
100 ~km s$^{-1}$.

The synchrotron emission from relativistic particles does not produces
a detectable X-ray flux.

\section{Comparison with \sax} \label{sec:sax}

While the numerical results and the analytic approximations derived
from them are primarily predictive, we can, at the very least,
directly apply them through comparisons to the one known XRB with a
bow shock nebula.

SAX J1712.6-3739 is so far the only XRB found to display a prominent
H$_{\alpha}$ bow shock nebula \citep{wie09}.  Following the original
discovery of the bow show, a deep VLT observation of the source was
obtained to confirm the detection and look for further structure in
the H-alpha map.  This observation is shown in Fig.~\ref{fig:sax}.

\subsection{Data reduction}

The data were taken using the Focal Reducer and low dispersion
spectrograph (FORS2) on the Very Large Telescope (VLT, Chile), using
the {\tt H$_\alpha$+83} interference filter. As part of a larger
programme (ESO programme ID 385.D-0100, PI Russell), we obtained 34
exposures of 60 seconds each on 16 April 2009, under excellent seeing
conditions (average 0.7 arcseconds).

The data were reduced using standard procedures in {\em IRAF} and
combined. Details on the reduction and flux calibration of these data
will be reported in a forthcoming publication (Russell et al. in
prep.).  The FORS2 data cover a much larger sky area than the EFOSC2
data reported in Wiersema et al. (2009) at a better resolution, and
have significantly better signal to noise, allowing us to search for
fainter features and both small and large scale structure.

\subsection{Model comparison}

\begin{figure}[!htbp]
  \centering
   \includegraphics[width=0.7\textwidth]{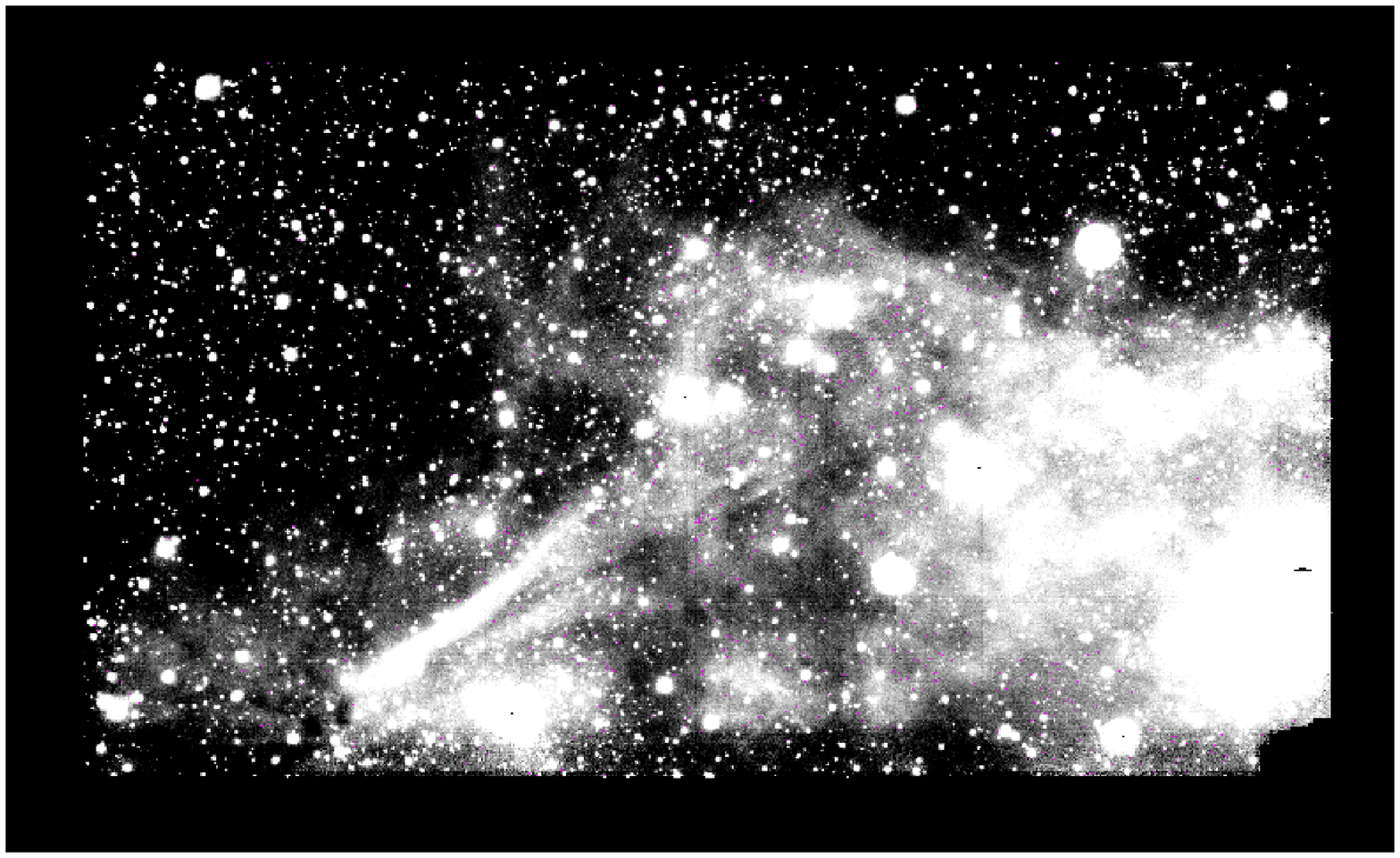}
   \includegraphics[width=0.7\textwidth]{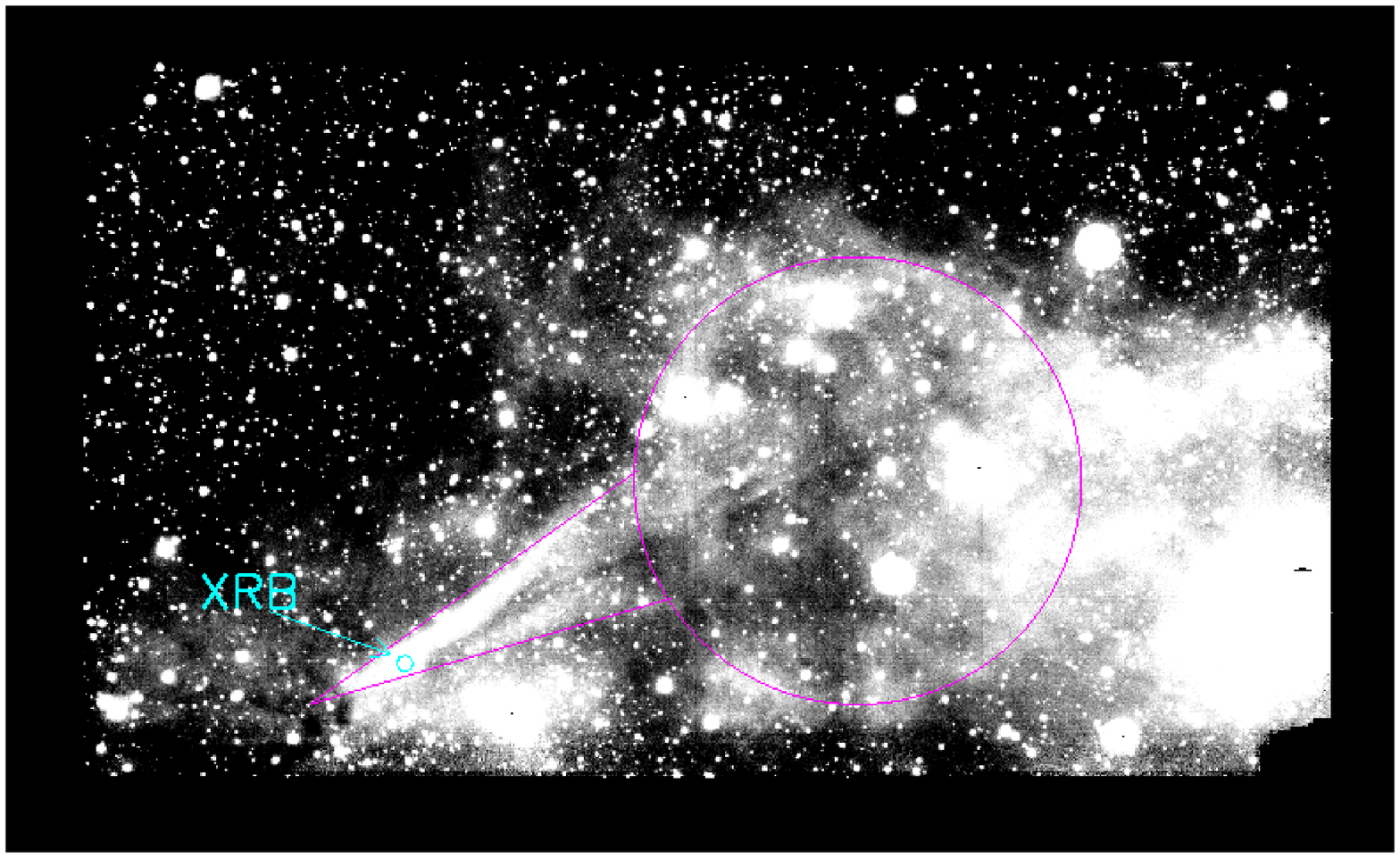}
   \caption{SAX J1712.6-3739 image from the VLT FORS2 H$_\alpha$ data
     (upper panel), and same image with location of the neck and
     bubble highlighted (lower panel).}
\label{fig:sax}
\end{figure}

The new VLT observation confirms the presence of a bright bow shock
nebula, as first reported in \citet{wie09}.  In addition, the image
suggests the presence of a roughly circular shell of H$_\alpha$
emission with a radius of $76$'' which connects to the previously
observed linear H$_\alpha$ features near the XRB.  While the level of
patchy background H$_{\alpha}$ emission makes a firm identification
difficult, we interpret this cavity or shell as the bubble inflated by
the backflow through the channel.  The estimated location of the
bubble and neck are sketched in the lower panel of figure
\ref{fig:sax}.  The distance from the end of the neck to the center of
the bubble is about $190$'' .

We estimate the half opening angle of the bow shock seen in the
H$_{\alpha}$ emission to be about $15^{\circ}$, which can be converted
into the projected oblique shock angle of $\beta = 0.261799$.  From
the observed ratio of bubble size to neck length, our analytic model
would predict a neck opening angle of about $19^{\circ}$, somewhat
larger than the observed value.  Assuming the bubble is expanding
supersonically, we can set a lower limit on the Mach number of the
source from eqn.~\ref{eqn:beta} of

\begin{eqnarray}\label{eqn:Ms}
  M_{s} \geq \frac{1+\beta}{\beta} M_{b} \geq 4.8
\end{eqnarray}

For our assumed sound speed of $17.3$ ~km s$^{-1}$, this implies the velocity
of the source is $v_s \geq 83$ ~km s$^{-1}$.  The source could also have a
component of motion towards or away from us which is not constrained
by the opening angle of the neck.

Given the distance of the object of $D\sim 7\,{\rm kpc}$ and the
approximate angular size of the putative shell of $76$'', we estimate
the physical radius of the bubble to be $R_{\rm bubble} \sim 2.6\,{\rm
  pc}$.  In order to check the consistency between the observation and
our model, we carried out a simulation with $v_s = 100$ ~km s$^{-1}$.  Using
the scaling relation of bubble radius from eqn.~(\ref{eqn:radi}) and a
standard ISM density of $1\,{\rm cm^{-3}}$, the appropriate luminosity
is $L = 8.5\times10^{33} ~\textrm{ergs s}^{-1}$ and the age of the
bubble would be approximately 70,000 years:
\begin{equation}\label{eqn:sax_age}
  t_{SAX} \approx 70,000 ~\textrm{yr}
  \left(\frac{L_{8.5e33}}{\rho_{0H}} \right)^{-1/3} 
  \left( \frac{R_{b}}{2.57 \textrm{pc}} \right)^{5/3}
\end{equation}
where $L_{8.5e33} = \frac{L}{8.5\times10^{33} ~\textrm{ergs s}^{-1}}$.

In terms of source power and ISM density, the velocity of \sax can be
estimates to be
\begin{eqnarray}\label{eqn:vs}
  v_{s} &=& 100 ~\textrm{km s}^{-1} \left(\frac{L_{8.5e33}}{\rho_{0H}}
  \right)^{-1/3}
\end{eqnarray}
A higher energy (or lower ambient density) implies a higher velocity required of the source by
eqn. \ref{eqn:vs} to have the same neck length at a given bubble size.  If the motion of the
binary is not perpendicular to our line of sight, but rather inclined at an angle $\alpha$, then
the true opening angle of the source is $\beta_{true} = \beta \sin{\alpha}$ and the velocity
of \sax would have to increase by $1/\sin{\alpha}$.  Eqn. \ref{eqn:vs} is therefore a lower
limit on the source velocity for a given set of parameters.  Based on the velocity analysis,
a proper motion of \sax can be estimated as a few mas/yr.  This proper motion scale will be
observable from the LSST. If the velocity is measured directly, it would help to constrain our
models and allow us to determine the jet power in terms of just the ambient density or vice versa.

In figure \ref{fig:sax}, the H$_\alpha$ emission in the neck is
brighter than that in the expanding bubble while we expect it to be
relatively dim from our fiducial simulations.

It is possible that this is due to the inclination angle for the
moving XRBs with respect to the line of sight. If the source is moving
at a substantial angle with respect to the plane of the sky, then the
amount of neck material we are looking through increases while the
thickness of the spherical bubble stays the same.  As a result, with
some large inclination angle, the emission of the neck can be brighter
than that of the bubble.

It is also possible that radiative cooling in the shell around the
bubble has moved much of the gas to temperatures too low to emit in
H$_{\alpha}$, while the neck is still at a sufficiently high
temperature to emit, given that the bow shock and Mach cone have a
higher Mach number and pressure than the bubble and shell.  Because
radiative cooling breaks the scale invariance of our simulations, a
more detailed investigation would require the construction of a large
grid of simulations at different source powers and ISM temperatures,
which would be well beyond the scope of this paper.

It is also possible that the relative brightness is simply a result of
the complicated morphology of background emission surrounding the
observed H$_{\alpha}$ nebula. The surrounding emission is clearly
non-uniform.  A more detailed and quantitative investigation of the
new VLT observations is forthcoming in a separate paper.

\begin{figure}[!htbp] 
  \begin{center}$
    \begin{array}{cc}
      \includegraphics[width=0.6\textwidth]{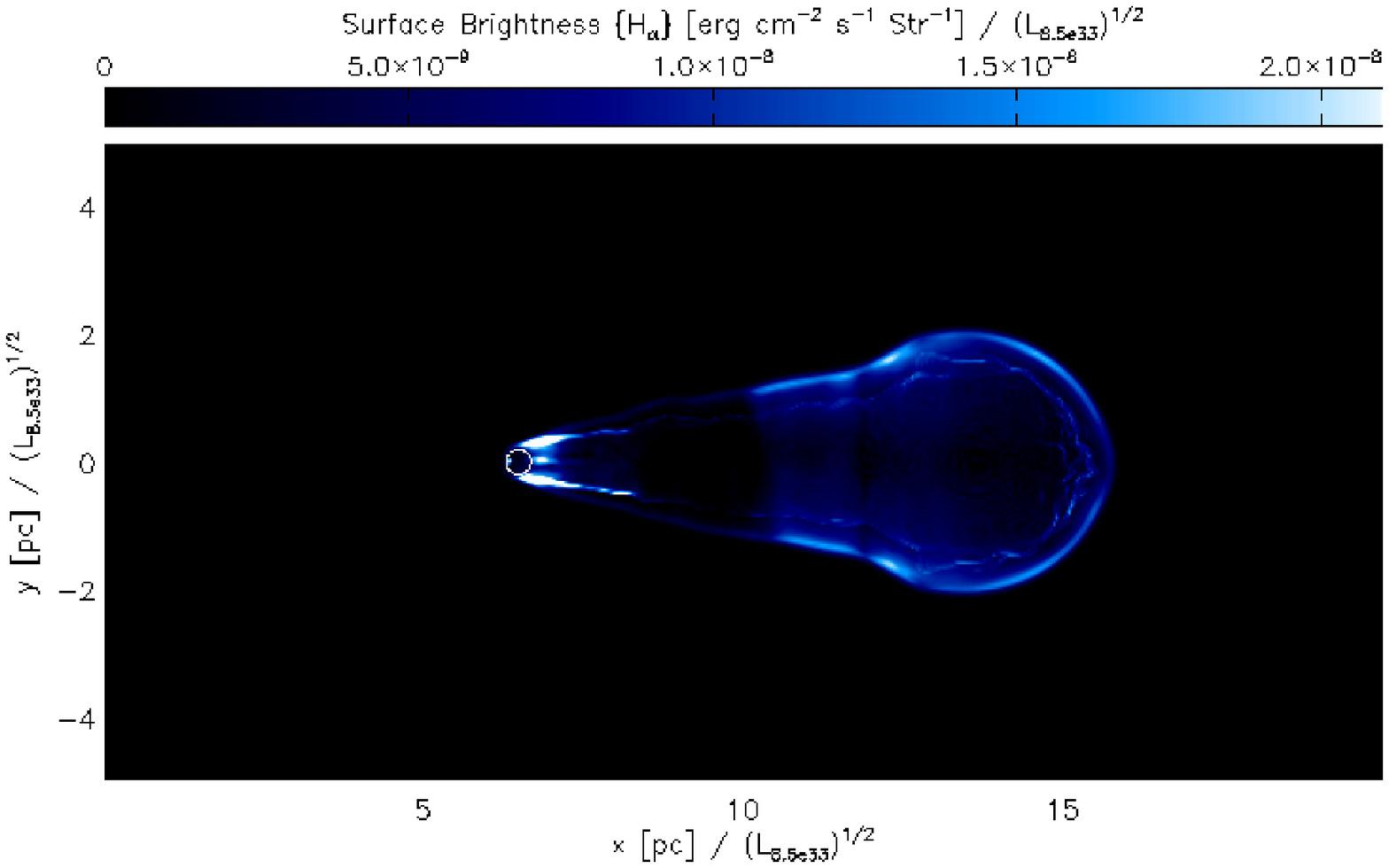} \\
      \includegraphics[width=0.6\textwidth]{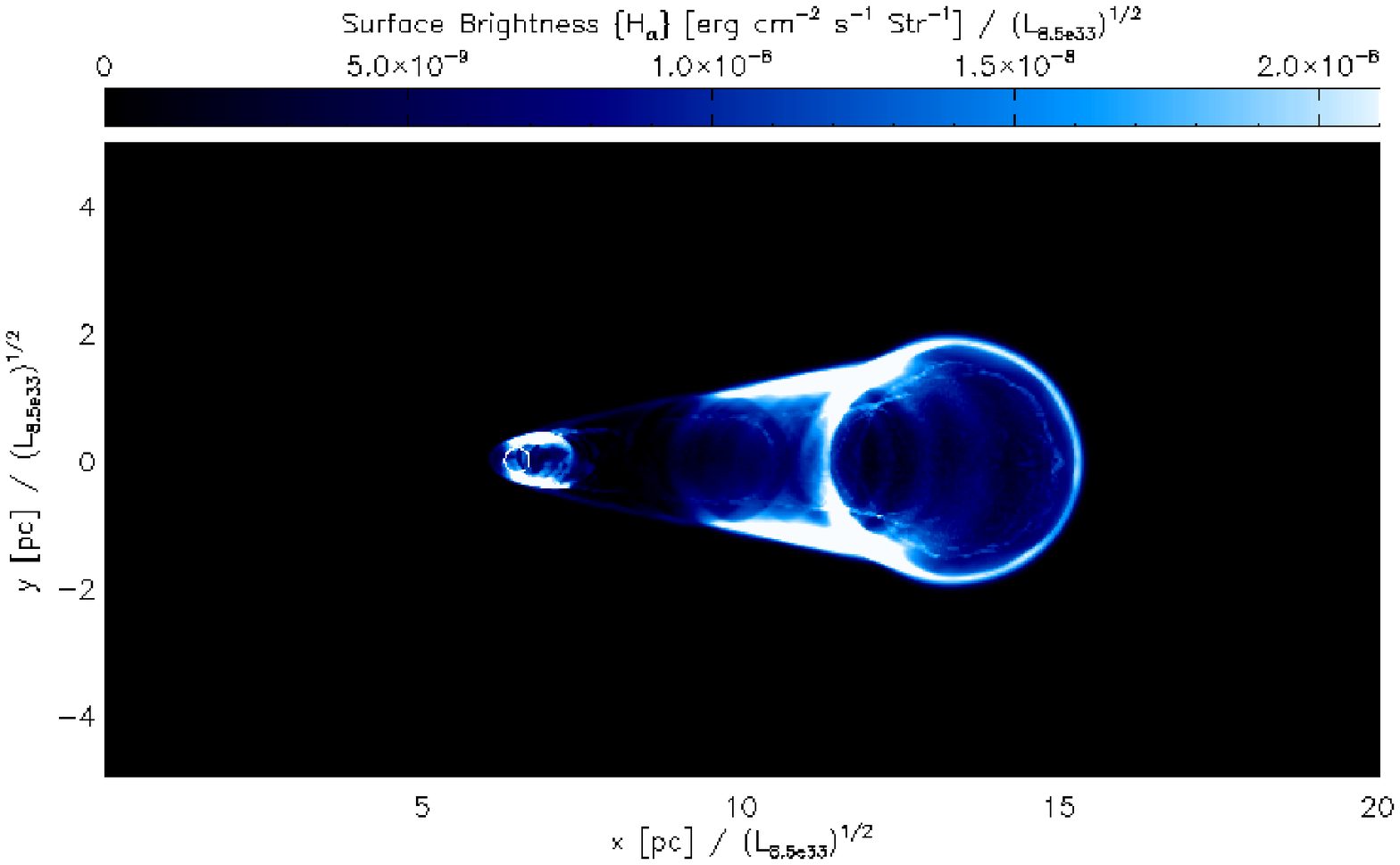} \\ 
qq    \includegraphics[width=0.6\textwidth]{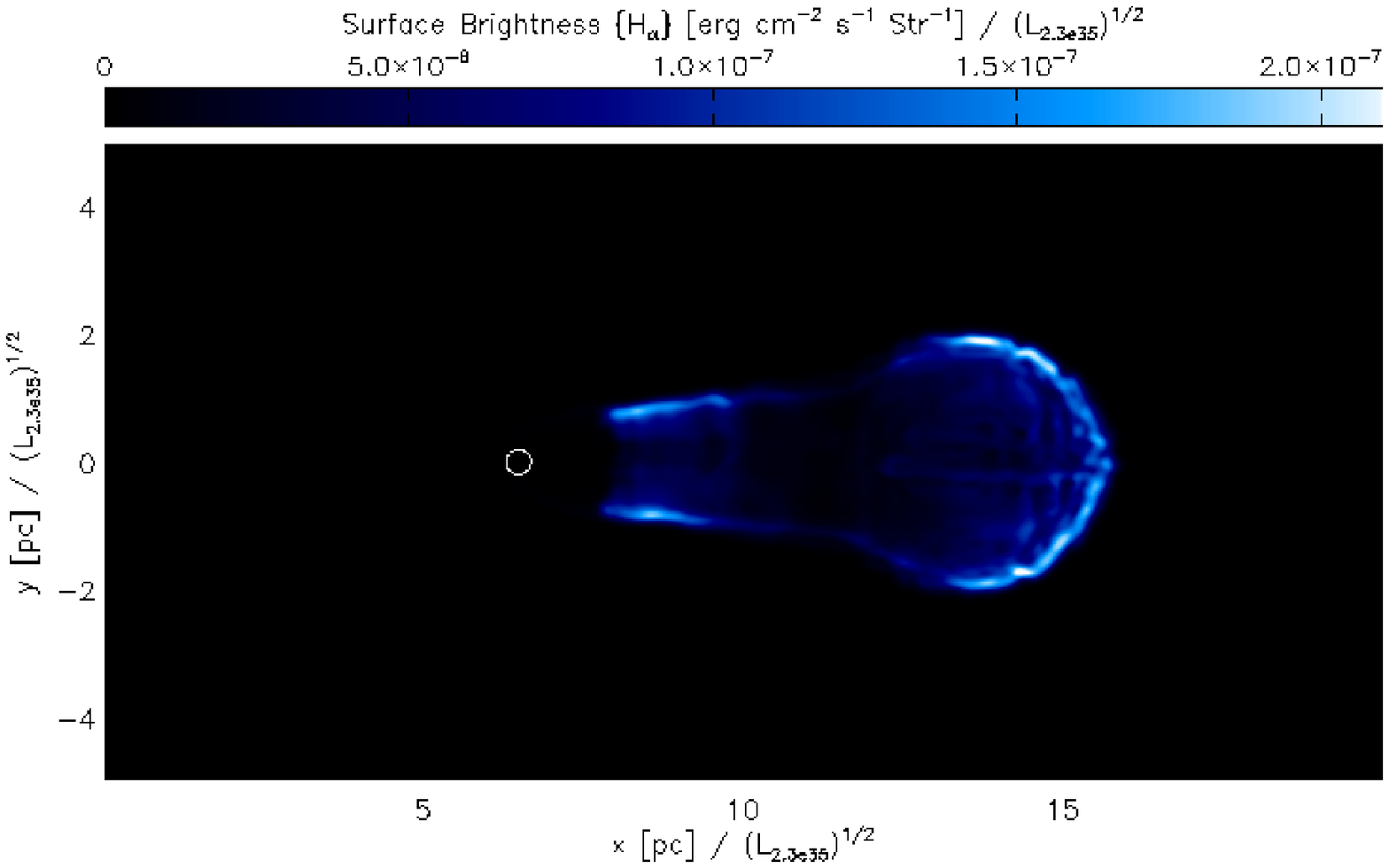}  
\end{array}$ 
\end{center}
\caption{H$_{\alpha}$ images with different initial conditions;
  $L=8.47\times10^{33} \textrm{ergs s}^{-1}$,~$v_{\rm XRB}=100
  ~\textrm{km s}^{-1}$ and edge-on at 69,000 yr (top panel),
  $L=8.47\times10^{33} \textrm{ergs s}^{-1}$,~$v_{\rm XRB}=200
  ~\textrm{km s}^{-1}$ and inclination angle of 60$^{\circ}$ at 69,000 yr
  (middle panel), $L=2.29\times10^{35} \textrm{ergs
    s}^{-1}$,~$v_{\rm XRB}=300 ~\textrm{km s}^{-1}$ and edge-on at 23,000 yr
  (bottom panel).  Note that the intensity scale of the bottom panel
  is 10 times higher than the other two panels.}
\label{fig:sax_sim} 
\end{figure}

The difficulty in interpreting the complex structure of the
observation notwithstanding, we carried out three simulations with the
initial conditions expected to produce a size and shape close to SAX
J1712.6-3739 to test whether we could match the observed shape and
relative brightness of the neck and bubble. The parameters of these
simulations are listed in table 1 as fit1, fit2 and fit3.  In figure
\ref{fig:sax_sim}, the top panel shows the results when the ejected
energy is $8.47\times10^{33} \textrm{ergs s}^{-1}$ and the velocity of
the XRB is $100$ ~km s$^{-1}$ (model fit1). There is bight H$_\alpha$ emission
in the bow shock and dimmer but still bright emission from the bubble.
However, the emission in most of the neck is relatively dim compared
to the shell.

In order to inspect the inclination angle effect, we ran simulation
fit2 with a higher velocity of $200$ ~km s$^{-1}$ and display it with an
inclination angle of 60$^{\circ}$ (middle panel).  In this case the
emission is brightest in the bow shock, part of the neck and the
leading edge of the bubble, with somewhat dimmer emission from the
trailing edge of the bubble.  This is somewhat closer to the bright
neck seen in SAX J1712.6-3739, although the location of bright neck
emission does not match the observation exactly.  The width of the
neck relative to the bubble size is about the same as the 100 ~km s$^{-1}$
case, and there is still a large gap in the neck with very little
emission.

Finally, we ran simulation fit3 with a larger power of
$L=2.29\times10^{35} \textrm{ergs s}^{-1}$, and higher velocity
velocity of $v_{\rm XRB}=300 ~\textrm{km s}^{-1}$ (bottom panel).  For the
given bubble size, the age is reduced to 23,000 years
(eqn.~\ref{eqn:sax_age}).  Note that because the emission is 
significantly brighter in this model, the intensity scale for the
bottom panel is increased by a factor of 10 compared to the other two
panels to avoid color saturation.  In this case, there is no
emission from the bow shock, due to the high temperature caused by the
faster source.  There is bright emission in part of the neck, where
the shock has become cool enough to emit in H$_\alpha$, and there is
about equally bright emission from the bubble edges.  By varying the
source velocity, it is possible to move the location of bright neck
emission from the bow shock (for model fit1) far down stream in the
neck (for model fit3) and it could be moved even farther away from the
XRB for a faster source velocity.

These three simulations differ substantially in their relative surface
brightness distributions of H$_{\alpha}$ emission in bow shock, neck,
and shell, showing that it should be possible to construct a
reasonable fit to the observations with sufficient fine tuning.

\section{Conclusion \& Summary}

By way of numerical simulations, we study the morphological evolution
and dynamics of microquasars with high space velocity, the functional
equivalent of narrow-angle tail radio galaxies for accreting stellar
mass black holes and neutron stars.  Such dynamic X-ray binaries
initially create a powered, spherically expanding shock/bubble, the
expansion of which eventually decelerates to less than the velocity of
the XRB relative to the ISM.

When the source reaches the edge of the bubble, it establishes a well
defined structure with a bow shock around the source at one end, a
spherical bubble at the other, and a neck connecting the bow shock and
bubble. The shock angle around neck, $\beta$, decreases as the binary
moves away from the bubble, but it cannot be less than 1/M$_0$, where
M$_0$ is the Mach number of the binary relative to the surrounding
medium.

The shocks from all three components creates H$_{\alpha}$ emission
that should be detectable in narrow band imaging.  At early times,
less than about 30,000 years in our simulations, X-ray emission will
be detectable around the bubble and neck, but this rapidly fades as
the shock temperature decreases.  At later times, X-ray emission is
only detectable from the bow shock and only for fast sources ($v_s
\geq 100$ ~km s$^{-1}$).

Synchrotron radio emission from the bubble is bright and covers a
large area, but emission from the neck is significantly weaker. There
is also strong synchrotron emission from near the binary, but it only
covers a small area and thus would be difficult to detect without high
resolution.  Bubbles created by galactic XRBs should be visible in
surveys of diffuse continuum emission. Binaries in other galaxies may
also be detectable with a resolution such that the beam size is
approximately the size of the bubble. The free-free emission is dimmer
than the synchrotron emission and is generally not detectable.

New H$_\alpha$ observations of SAX J1712.6-3739 presented here (see
\S\ref{sec:sax}) show both a strong bow shock and tentative evidence
for a spherical shell trailing the XRB, consistent with the neck and
bubble morphology predicted by our simulations.  We are also able to
constrain the Mach number of the XRB relative to the background ISM to
be $M_s \geq 4.8$, corresponding to a velocity of $v_s \geq 83$ ~km s$^{-1}$
in the plane of the sky for a sound speed of $17.3$ ~km s$^{-1}$.  

Extending our results into pulsar bow shock nebulae (PWBN) is plausible, 
since bow shock structures of PWBN are well described by the simulations 
presented in this paper. For example, the H$_\alpha$ images 
of PSR B2224+65 \citep{cha02,cor93},
called ``Guitar nebula'' for its peculiar shape, shows well developed 
structures of trailing neck and spherical bubble.
We will present models specific to PWBN in a future paper.

\acknowledgments{
We thank Hand-Jakob Grimm, Ellen Zweibel, and Brian Gaensler for helpful discussions. 
S.H. and D.Y. acknowledges support from NASA/Chandra Theory grant TM9-002X. 
S.H., D.Y., and B.M. acknowledges support through NSF grant AST-0908690.
K.W. acknowledge support from STFC. Based in part on observations made with ESO Telescopes at
the Paranal Observatory under programme ID 385.D-0100
}

\bibliographystyle{apj}
\bibliography{xrbs_wind}

\end{document}